\documentclass[aps,prb,twocolumn,amsmath,amssymb,floatfix,groupedaddress]{revtex4-1}
\usepackage{graphicx}
\begin{document}
\title{Measuring Which-Path Information with \\
Coupled Electronic Mach-Zehnder Interferometers}
\author{J. Dressel, Y. Choi, A. N. Jordan}
\affiliation{Department of Physics and Astronomy, University of Rochester, Rochester, New York 14627, USA}
\date{\today}

\newcommand{\mean}[1]{\langle #1 \rangle}
\newcommand{\cmean}[2]{\,_{#1}\langle #2 \rangle}
\newcommand{\ket}[1]{|#1\rangle}
\newcommand{\bra}[1]{\langle #1|}
\newcommand{\iprod}[2]{\langle #1 | #2 \rangle}
\newcommand{\oprod}[2]{\ket{#1}\bra{#2}}
\newcommand{\hsp}[2]{\left\langle #1, #2\right\rangle}
\newcommand{\prj}[1]{\oprod{#1}{#1}}
\newcommand{\norm}[1]{\iprod{#1}{#1}}
\newcommand{\expect}[2]{\mean{#2}_{#1}}
\newcommand{\wexpect}[3]{_{#1}\mean{#3}_{#2}}
\newcommand{\op}[1]{\hat{#1}}
\newcommand{\cre}[1]{\op{a}^\dagger_{#1}}
\newcommand{\ann}[1]{\op{a}_{#1}}
\newcommand{\Tr}[1]{\text{Tr}\left[#1\right]}

\begin{abstract}
  We theoretically investigate a generalized ``which-path'' measurement on an electronic Mach-Zehnder Interferometer (MZI) implemented via Coulomb coupling to a second electronic MZI acting as a detector.  The use of contextual values, or generalized eigenvalues, enables the precise construction of which-path operator averages that are valid for any measurement strength from the available drain currents.  The form of the contextual values provides direct physical insight about the measurement being performed, providing information about the correlation strength between system and detector, the measurement inefficiency, and the proper background removal.  We find that the detector interferometer must display maximal wave-like behavior to optimally measure the particle-like which-path information in the system interferometer, demonstrating wave-particle complementarity between the system and detector.  We also find that the degree of quantum erasure that can be achieved by conditioning on a specific detector drain is directly related to the ambiguity of the measurement.  Finally, conditioning the which-path averages on a particular system drain using the zero frequency cross-correlations produces conditioned averages that can become anomalously large due to quantum interference; the weak coupling limit of these conditioned averages can produce both weak values and detector-dependent semi-weak values.
\end{abstract}

\maketitle

\section{Introduction}
The construction of electronic Mach-Zehnder interferometers (MZIs) in the solid state is a recent innovation in the fabrication and control of coherent mesoscopic systems.  The first experiment of this kind, published by the Heiblum group, \cite{Ji2003} used the edge states~\cite{Devyatov2007} of an integer quantum Hall Corbino geometry as the electronic analog of light beams and quantum point contacts (QPCs) \cite{Buttiker1992,Martin1992,Blanter2000} as the electronic analogs of optical beam splitters to construct an interferometer with a visibility of 62\%.  Other interferometer designs have since been similarly implemented as electronic interferometers in the integer quantum Hall regime. \cite{Neder2007b,Giovannetti2008,Lin2009,Chirolli2010}

The electronic MZI differs from its optical counterpart in several respects.  The arms of the MZI accumulate relative phase differences not only due to kinetic propagation of electrons along the arms, but also because the electrons are charged particles and can thus acquire a geometric Aharanov-Bohm phase\cite{Aharonov1959,Anandan1992} when the arms enclose a magnetic flux.  This charge can also lead to strong electron-electron interactions, giving rise to a variety of effects that have no counterpart in an optical MZI.  For example, the interactions can produce differences in the counting statistics, \cite{Chung2005,Forster2005} can induce temperature-dependent decoherence, \cite{Chalker2007,Litvin2007,Neder2007a,Roulleau2008,Youn2009,Hashisaka2010} can be used to detect external charges, \cite{Khym2009,Haack2010} and can even lead to lobe structure in the visibility at high voltage bias.\cite{Neder2006,Roulleau2007,Litvin2008,Bieri2009}  Such lobe structure was unexpected and has generated numerous theoretical explanations, \cite{Sukhorukov2007a,Levkivskyi2008,Neder2008,Neuenhahn2008,Youn2008,Levkivskyi2009b,Kovrizhin2009,Kovrizhin2010} some hypothesizing Luttinger liquid physics as the cause.  

\begin{figure}[b]
  \begin{center}
    \includegraphics[width=8cm]{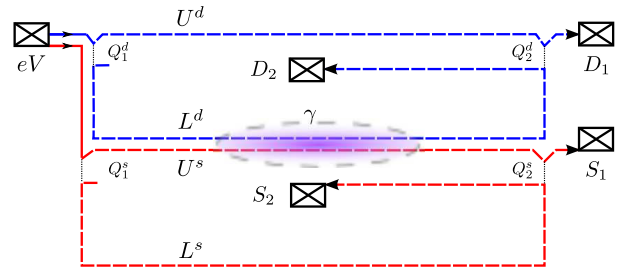}
  \end{center}
  \caption{(color online) Schematic of coupled electronic MZIs.  An ohmic source in a quantum Hall system at filling factor $\nu=2$ injects chiral excitation pairs biased at energy $eV$ relative to the ohmic reference drains $D_1$, $D_2$, $S_1$, and $S_2$ into independent edge channels.  The bias is kept low enough to allow only one excitation per channel on average.  The outer (red) channel is transmitted entirely through the QPC $Q^d_1$ and partially transmitted through $Q^s_1$ and $Q^s_2$, forming the system MZI.  The inner (blue) channel is reflected entirely from $Q^s_1$ and $Q^s_2$ and partially transmitted through $Q^d_1$ and $Q^d_2$, forming a separate detector MZI.  The Coulomb interaction between the copropagating arms $L^d$ and $U^s$ induces an average relative phase shift $\gamma$ between each excitation pair that couples the interferometers.}
  \label{fig:MZI}
\end{figure}

Here we take a more modest theoretical approach for describing electronic MZIs in a quantum Hall system that focuses on the low bias regime within a single-particle edge-state model.  We consider a configuration of two such single-particle electronic MZIs coupled together by the Coulomb interaction, as shown in Figure~\ref{fig:MZI}.  This geometry has similarities to Hardy's paradox \cite{Hardy1992,Aharonov2001,Lundeen2009,Yokota2009} and has been considered previously at various levels of detail by several authors \cite{Kang2007,Kang2008,Braun2008,Shpitalnik2008} to explore such phenomena as quantum erasure \cite{Wheeler1979,Scully1982,Scully1991,Herzog1995,Mir2007,Hilmer2007} and Bell inequality violations.  In our treatment, we include interactions between the MZIs via a minimal phenomenological model that adds a relative interaction-induced phase shift between a pair of electrons that occupy adjacent edge states simultaneously.  The relative phase shift has the effect of entangling \cite{Wootters1998,Samuelsson2005,Frustaglia2009} the states, mixing the path information of the two MZIs.  This kind of interaction has been experimentally shown to be capable of producing a $\pi$-phase shift on a single electron---perhaps the most dramatic difference from the optical analog. \cite{Neder2007a}  Thus, all elements of our theoretical analysis are based on currently available technology.

Our work considers the task of detecting which-path information in one MZI by using the second MZI as the detector.  Since the system and the detector are identical devices, this arrangement has several appealing features.  First, the symmetry of the geometry indicates that there should be a duality between ``which-path'' information in one MZI versus ``which-fringe'' information in the other.  We will show that this is indeed the case, which relates this work to earlier ``controlled dephasing'' experiments. \cite{Buks1998,Sprinzak2000,Neder2007c,Roulleau2009}  We apply the contextual values formalism \cite{Dressel2010,Dressel2011} for generalized measurements \cite{Braginski1992,Nielsen2000,Wiseman2009} to show how even with inefficient detection \cite{Buttiker2002,Averin2005} and low visibility, the which-path information may be extracted from the detector currents systematically.  Next, the fact that both the system and the detector have their own inputs, outputs and coherence allows the effects of measurement to be explored in detail.  In particular, the correlations between them can be experimentally measured and analyzed, taking various forms such as joint counting statistics, or even conditioned measurements (see Sukhorukov et al.\cite{Sukhorukov2007b} for an example with incoherent electrons).  The ability to condition (or post-select) measurements performed on a system with quantum coherence also allows the possibility of measuring weak values. \cite{Aharonov1988,Ritchie1991,Pryde2005,Aharonov2008,Dressel2010}  Weak values, in addition to being of interest in their own right, have been shown to be useful as an amplification technique for measuring small variations of a system parameter \cite{Hosten2008,Dixon2009,Starling2009,Howell2010,Zilberberg2011}, as well as for tests of bona fide quantum behavior.  \cite{Leggett1985,Williams2008,Dressel2011,Aharonov2001,Tollaksen2007}

The paper is organized as follows.  In Section~\ref{sec:geometry}, we describe the considered geometry and our physical modeling principles before reviewing and applying the scattering theory of mesoscopic transport to the two edge state physics.  In Section~\ref{sec:measurement}, we take the joint predictions of the scattering model and interpret them as a measurement by the detector electron that extracts information about the path state of the system electron.  We describe a principled way to construct this information from the data in a variety of regimes.   In Section~\ref{sec:conditioning} we introduce the conditioning procedure and use it to clarify the phenomenon of quantum erasure, as well as to calculate conditioned averages of the which-path measurements.  We shall see that both weak values and semi-weak values will compete as weak coupling limits of these conditioned averages.  In Section~\ref{sec:conclusion} we describe our conclusions.

\section{Coupled MZIs}\label{sec:geometry}
We consider a pair of electronic MZIs embedded in a two-dimensional electron gas in the integer quantum Hall regime at filling factor $\nu=2$ as illustrated in Figure~\ref{fig:MZI}.  An ohmic source with a small DC-bias $eV$ above the Fermi energy $E_F$ injects chiral electron-like edge excitations of charge $e$ into the sample that propagate uni-directionally along two independent edge channels.~\cite{Devyatov2007}  Each edge channel forms an interferometer from two appropriately tuned quantum point contacts (QPCs) that coherently split and then recombine the possible paths.  The relative phase between the arms of each interferometer is determined not only by a local dynamical phase accumulated during kinetic propagation along each arm, but also by a global geometric phase~\cite{Anandan1992} (in the form of the Aharonov-Bohm (AB) effect) \cite{Aharonov1959} arising from the closed paths.  After the paths interfere, the charges are collected at ohmic reference drains held at the Fermi energy, producing fluctuating output currents that can be temporally averaged.

The two interferometers accrue an additional relative phase shift due to the Coulomb interaction where the charges copropagate.  Intuitively, the mutual repulsion affects the dynamical phases by effectively warping the propagation paths, which also affects the geometrical phases by changing the areas enclosed by the paths.  A more careful analysis of the joint interaction phase is provided in the Appendix.  Such additional relative phase has the effect of \emph{entangling} the joint state of the two interferometers, mixing the which-path information.  Due to the entanglement, extracting information from the drains of one interferometer allows one to infer correlated which-path information about the other interferometer.  That is, one interferometer can be used as a detector to \emph{indirectly measure}~\cite{Dressel2010,Braginski1992,Nielsen2000,Wiseman2009} the which-path information of the other.  As we shall see, the characteristics of the measurement will depend on the tuning of the detector interferometer as well as on the coupling phase.

We model the coupled MZI system using the elastic scattering approach of Landauer and B\"uttiker~\cite{Buttiker1992,Martin1992,Blanter2000} for coherent charge transport.  As the transport is largely ballistic in the integer quantum Hall regime, the formalism directly relates the average currents $I_l$ collected at each ohmic lead $l\in\{D_1, D_2, S_1, S_2\}$ to the \emph{transmission probabilities} $P_l(E)\in[0,1]$ for plane waves of fixed energy $E$ to traverse the sample successfully.  Treating the ohmic leads as thermal reservoirs, the average currents from spinless single-channel transport are,
\begin{subequations}\label{eq:currents}
\begin{align}
  I_l &= e \int_{0}^{eV} \frac{dE}{h} (f(E + eV) - f(E)) P_l(E) \\
  \label{eq:current_approx}
  &\approx \frac{e^2 V}{h} P_l(E_F).
\end{align}
\end{subequations}
Here, $f(E) = (\exp\left( (E - E_F)/k_B T \right) + 1)^{-1}$ is the equilibrium Fermi distribution relative to the Fermi energy $E_F$ at a temperature $T$; $h$ is Planck's constant; and, $k_B$ is Boltzmann's constant.  

The approximate equality \eqref{eq:current_approx} holds in the low-bias regime when $E_F \gg eV \gg k_B T$ and the transmission probabilities $P_l$ are constant with respect to the small variations in energy.  We assume that the source operates in such a regime.  Due to the small spectral width of the source, the fermionic excitations will then be well approximated as plane waves at a fixed energy on the scale of the sample; hence, on average only one excitation per channel will occupy the sample and intrachannel interactions can be ignored.  In particular, we avoid the anomalous lobe structure in the interference that appears at higher bias. \cite{Neder2006,Roulleau2007,Litvin2008,Bieri2009}

We also assume for simplicity of discussion that the source only injects spinless excitation \emph{pairs} with one excitation per channel so that the coupling interaction between the channels is fixed; the results will be averaged over a more realistic source distribution in Section~\ref{sec:fluctuation}.  With these approximations, the initial joint scattering state for an excitation pair can be written in second-quantized notation as,
\begin{equation}\label{eq:initstate}
  \ket{\Psi} = \op{a}^{d \dagger}\op{a}^{s \dagger}\ket{0},
\end{equation}
where $\ket{0}$ is the filled Fermi sea of the edge channels and $\op{a}^{d \dagger}$ and $\op{a}^{s \dagger}$ are creation operators for plane waves of a fixed energy injected into the inner and outer channels, respectively.  Operators corresponding to different edge channels commute due to the independence of the channels.  

The inner channel will form a Mach-Zehnder interferometer as shown in Figure~\ref{fig:MZI}, which we refer to as the upper MZI, or the \emph{detector} MZI.  Similarly, the outer channel will form an identical interferometer, which we refer to as the lower MZI, or the \emph{system} MZI.  We will use the lowercase superscripts $^d$ and $^s$ throughout to distinguish quantities specific to the detector and the system, respectively, and to avoid confusion with the detector and system drains that we denote with capital letters $D_1$, $D_2$, $S_1$ and $S_2$.

The QPCs $Q^d_1$, $Q^d_2$, $Q^s_1$, and $Q^s_2$ shown in Figure~\ref{fig:MZI} each elastically scatter the plane waves, affecting only the complex amplitudes of the joint scattering state.  Hence, for $m\in\{d,s\}$, $i\in\{1,2\}$ we can represent the effect of each QPC as a unitary scattering matrix, 
\begin{align}\label{eq:qpcs}
  \op{U}^m_i &= \begin{pmatrix}e^{i \chi^m_i} t^m_i\,  & e^{i \xi^m_i} r^m_i\,  \\
                           e^{i \chi^m_i} r^m_i & e^{i \xi^m_i} t^m_i \end{pmatrix},
\end{align}
where $t^m_i = \sqrt{T^m_i}$ and $r^m_i = i \sqrt{R^m_i}$ are given in terms of the transmission and reflection probabilities $T^m_i \in[0,1]$ and $R^m_i =1-T^m_i$ though $Q^m_i$.  The additional scattering phases $\chi^m_i$ and $\xi^m_i$ may arise from QPC asymmetry.

The QPCs are kept tunable subject to the constraints that the outer channel is fully transmitted through $Q^d_1$ and $Q^d_2$ and the inner channel is fully reflected from $Q^s_1$ and $Q^s_2$ to create the two separate interfering paths.  There is an additional QPC near drain $S_1$ not shown in Figure~\ref{fig:MZI} that is kept fixed to allow full transmission of the outer channel and full reflection of the inner channel in order to divert the outer channel for collection at the drain $S_1$.  

\begin{figure}[t]
  \begin{center}
    \includegraphics[width=\columnwidth]{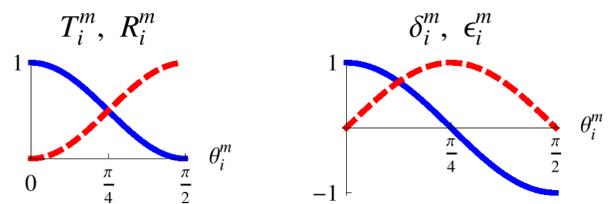}
  \end{center}
  \caption{(color online) Complementary QPC balance parameters \eqref{eq:qpcbalance} for $m\in\{d,s\}$ and $i\in\{1,2\}$ as parametrized by the balance angle $\theta^m_i$.  Left: transmission probability $T^m_i$ (solid, blue) and reflection probability $R^m_i$ (dashed, red).  Right: particle-like parameter $\delta^m_i$ (solid, blue) and wave-like parameter $\epsilon^m_i$ (dashed, red).}
  \label{fig:qpcparams}
\end{figure}

For later convenience we also introduce the complementary QPC balance parameters,
\begin{subequations}\label{eq:qpcbalance}
\begin{align}
  \label{eq:delta}
  \delta^m_i &= T^m_i - R^m_i  \in [-1,1], \\
  \label{eq:epsilon}
  \epsilon^m_i &= 2\sqrt{T^m_i\, R^m_i} \in [0,1], 
\end{align}
\end{subequations}
for $m\in\{d,s\}$ and $i\in\{1,2\}$ that satisfy $(\epsilon^m_i)^2 + (\delta^m_i)^2 = 1$.  All such QPC parameters can be related by a QPC balance angle $\theta^m_i \in[0,\pi/2]$ such that, $T^m_i = \cos^2\theta^m_i$, $R^m_i = \sin^2\theta^m_i$, $\delta^m_i = \cos2\theta^m_i$, and $\epsilon^m_i = |\sin2\theta^m_i|$ as illustrated in Figure~\ref{fig:qpcparams}.  We shall see that the parameters $\delta^m_i$ control the particle-like path-bias of the excitation after a QPC, while the parameters $\epsilon^m_i$ control the complementary wave-like interference visibility.

The joint state \eqref{eq:initstate} can be scattered through $Q^d_1$ and $Q^s_1$ using \eqref{eq:qpcs} into the basis of the MZI paths, yielding the replacements,
\begin{subequations}
\begin{align}
  \op{a}^{d \dagger} &= e^{i \chi^d_1\, } t^d_1\, \cre{L^d} + e^{i \xi^d_1\, } r^d_1\, \cre{U^d}, \\ 
  \op{a}^{s \dagger} &= e^{i \chi^s_1\, } t^s_1\, \cre{L^s} + e^{i \xi^s_1\, } r^s_1\, \cre{U^s}.
\end{align}
\end{subequations}
During propagation to the second pair of QPCs, each path $p\in\{L^d, U^d, L^s, U^s\}$ accumulates an additional dynamical phase $\phi_p$ that depends on the excitation energy and the path-length.  When the paths recombine, the difference between the dynamical phases contributes to the interference.  Closing the paths for MZI $m\in\{s,d\}$ also contributes a relative geometric Aharonov-Bohm (AB) phase $\phi^m_{AB}$ that depends on the magnetic flux enclosed by the path.  

We can compactly account for the various phase effects contributing to the interference by defining \emph{tuning phases} for each MZI,
\begin{subequations}\label{eq:tuning}
\begin{align}
  \phi^d &= \phi^d_{AB} + \phi_{L^d} - \phi_{U^d} + \chi^d_1\,  - \xi^d_1\, , \\
  \phi^s &= \phi^s_{AB} + \phi_{L^s} - \phi_{U^s} + \chi^s_1\,  - \xi^s_1\, .
\end{align}
\end{subequations}

Finally, the joint scattering amplitude corresponding to co-occupation of $L^d$ and $U^s$ acquires a effective Coulomb interaction phase $\gamma$ that couples the two interferometers.  (See the Appendix for discussion about how the Coulomb effect can produce such a phase shift.)  This interaction phase compactly encodes opposing shifts in the combined dynamical and geometric phases of each MZI due to the Coulomb repulsion of the charge pair.  For simplicity, we assume for now that the relative phase is constant; we will allow it to fluctuate for consecutive pairs in Section \ref{sec:fluctuation}.  We also note that any additional Coulomb phase acquired during copropagation after QPC $Q^d_1$ and before QPC $Q^s_1$ will only contribute to the tuning phase $\phi^d$ and can therefore be ignored.

After adding the phenomenological phases, the scattered joint state just before the second pair of QPCs is,
\begin{align}\label{eq:scatterstatearms}
  \ket{\Psi'} &= \Big(t^d_1\, t^s_1\, e^{i (\phi^d + \phi^s)} \cre{L^d}\cre{L^s} + r^d_1\, r^s_1\, \cre{U^d}\cre{U^s} \\
  & \;{} + r^d_1\, t^s_1\, e^{i \phi^s}\cre{U^d}\cre{L^s} + t^d_1\, r^s_1\, e^{i (\phi^d + \gamma)}\cre{L^d}\cre{U^s}\Big)\ket{0}, \nonumber
\end{align}
up to a global phase of $\exp(i(\phi_{U^d}+\phi_{U^s}+\xi^d_1 +\xi^s_1))$ not written.  

The interaction phase $\gamma$ has the effect of \emph{entangling} the two interferometers, which we can show by computing the concurrence,~\cite{Wootters1998} 
\begin{align}
  \mathcal{C}\Big[\ket{\Psi'}\Big] &= \epsilon^d_1\, \epsilon^s_1\,  \left|\sin\frac{\gamma}{2}\right| \in [0,1].
\end{align}
We see that the entanglement reaches a maximum when the phase $\gamma\to\pi$ and vanishes as $\gamma\to 0$.  Furthermore, the entanglement directly depends on the QPCs $Q^d_1$ and $Q^s_1$ preparing interfering wave-like excitations in each MZI, which is measured by $\epsilon^d_1\, \epsilon^s_1\, $; maximum entanglement can only occur for balanced QPCs with $T^d_1\,  = T^s_1\,  = 1/2$, or $\epsilon^d_1\, \epsilon^s_1\, =1$.

At this point, we conceptually break the symmetry between the two interferometers to treat one as a detector for information about the other.  We will treat the upper MZI as the \emph{detector} and the lower MZI as the \emph{system} being measured, though obviously we could exchange those roles by the symmetry of the geometry.  To do this we finish scattering the detector MZI through $Q^d_2$ into the basis of the ohmic detector drains $\{D_1,D_2\}$ using \eqref{eq:qpcs},
\begin{align}
  \begin{pmatrix}\cre{L^d}\\ \cre{U^d}\end{pmatrix} &= \op{U}^d_2 \begin{pmatrix}\cre{D_1} \\ \cre{D_2}\end{pmatrix}
\end{align}
yielding,
\begin{align}
  \label{eq:scatterstate}
  \ket{\Psi''} &= \Big(\cre{D_1}\left[ C_{D_1,L^s}e^{i \phi^s}t^s_1\, \cre{L^s} + C_{D_1,U^s}r^s_1\, \cre{U^s}\right] \\
  &\quad {} + \cre{D_2}[ C_{D_2,L^s}e^{i \phi^s}t^s_1\, \cre{L^s} + C_{D_2,U^s}r^s_1\, \cre{U^s}]\Big)\ket{0}, \nonumber
\end{align}
up to the same global phase as in \eqref{eq:scatterstatearms}.  For later convenience, we have defined the detector scattering amplitudes,
\begin{subequations}\label{eq:scattering}
\begin{align}
  C_{D_1,L^s} &= e^{i \chi^d_2\, } \left(t^d_1\, t^d_2\, e^{i \phi^d} + r^d_1\, r^d_2\, \right), \displaybreak[0] \\
  C_{D_1,U^s} &= e^{i \chi^d_2\, } \left(t^d_1\, t^d_2\, e^{i (\phi^d + \gamma)} + r^d_1\, r^d_2\, \right), \displaybreak[0] \\
  C_{D_2,L^s} &= e^{i \xi^d_2\, } \left(t^d_1\, r^d_2\, e^{i \phi^d} + r^d_1\, t^d_2\, \right), \displaybreak[0] \\
  C_{D_2,U^s} &= e^{i \xi^d_2\, } \left(t^d_1\, r^d_2\, e^{i (\phi^d + \gamma)} + r^d_1\, t^d_2\, \right).
\end{align}
\end{subequations}

\section{Measurement Interpretation}\label{sec:measurement}
The joint scattering model is useful for computing probabilities and average currents, but it does not provide direct insight into the measurement being performed by one interferometer on the other.  To make the connection to measurement more apparent, we will use the \emph{contextual values} formalism~\cite{Dressel2010,Dressel2011} that links the detector drain probabilities directly to the ``which-path'' operator for the system.  We will see that we can understand the various subtleties of the measurement quite transparently using this technique.

\subsection{POVM}
To facilitate the interpretation of the distinguishable detector drains as the \emph{outcomes} of a measurement being performed on the system, we define the single particle state kets,
\begin{subequations}
\begin{align}
  \ket{D_1} &= \cre{D_1}\ket{0}, & \ket{D_2} &= \cre{D_2}\ket{0}, \\
  \ket{L^s} &= \cre{L^s}\ket{0}, & \ket{U^s} &= \cre{U^s}\ket{0},
\end{align}
\end{subequations}
define the reduced system state in absence of interaction,
\begin{align}\label{eq:reducedstate}
  \ket{\psi^s} &= e^{i \phi^s} t^s_1\, \ket{L^s} + r^s_1\, \ket{U^s},
\end{align}
and write \eqref{eq:scatterstate} in the form,
\begin{align}\label{eq:scattermeasure}
  \ket{\Psi''} &= \ket{D_1}\otimes\op{M}_{D_1}\ket{\psi^s} + \ket{D_2}\otimes\op{M}_{D_2}\ket{\psi^s}.
\end{align}

The interaction with the detector in \eqref{eq:scattermeasure} is entirely represented by operators acting on the reduced \emph{system} state \eqref{eq:reducedstate} that contain all the scattering information of the detector,
\begin{subequations}\label{eq:measops}
\begin{align}
  \op{M}_{D_1} &= C_{D_1,L^s} \prj{L^s} + C_{D_1,U^s} \prj{U^s}, \\
  \op{M}_{D_2} &= C_{D_2,L^s} \prj{L^s} + C_{D_2,U^s} \prj{U^s}.
\end{align}
\end{subequations}
The operator $\op{M}_{D_1}$ encodes the interaction followed by the absorption of the detector excitation at the drain $D_1$.  Similarly, the operator $\op{M}_{D_2}$ encodes the interaction followed by absorption at $D_2$.  We refer to $\{\op{M}_{D_1},\op{M}_{D_2}\}$ as \emph{measurement operators}.~\cite{Braginski1992,Nielsen2000,Wiseman2009}

As the coupling phase $\gamma\to0$ the measurement operators \eqref{eq:measops} become nearly proportional to the identity.  We call $\gamma\to0$ the \emph{weak coupling limit} since the reduced system state is only weakly perturbed for small $\gamma$.  Conversely the limit $\gamma\to\pi$ is called the \emph{strong coupling limit} since the measurement operators are maximally different from the identity and maximally perturb the reduced system state.

The measurement operators also form a \emph{Positive Operator-Valued Measure} (POVM) on the system,
\begin{align}\label{eq:povm}
  \op{E}_{D_1} &= \op{M}_{D_1}^\dagger \op{M}_{D_1}, & \op{E}_{D_2} &= \op{M}_{D_2}^\dagger \op{M}_{D_2},
\end{align}
such that $\op{E}_{D_1} + \op{E}_{D_2} = \op{1}$.  The POVM elements $\{\op{E}_{D_1},\op{E}_{D_2}\}$ act as \emph{probability operators} for the measurement outcomes.

Hence, the probability of absorbing the detector excitation at a drain $D\in\{D_1,D_2\}$ can be expressed either as an expectation of the projection operator of the detector drain under the joint state \eqref{eq:scatterstate} \emph{or}, equivalently, as an expectation of the probability operator \eqref{eq:povm} under the unperturbed system state \eqref{eq:reducedstate},
\begin{align}\label{eq:probs}
  P_{D} &= |\iprod{D}{\Psi''}|^2 = \bra{\psi^s}\op{E}_{D}\ket{\psi^s} \\
         &= |C_{D,L^s} t^d_2\, |^2 + |C_{D,U^s} r^d_2\, |^2. \nonumber 
\end{align}

By working with the reduced state \eqref{eq:reducedstate}, the measurement operators \eqref{eq:measops}, and the probability operators \eqref{eq:povm}, we treat the detector as an abstract entity whose sole purpose is to measure the system.  Such abstraction allows us to more clearly examine the measurement being made upon the system.

\subsection{Contextual Values}
In order to relate the measurement on the system to observable information that we can interpret, we will use \emph{contextual values}~\cite{Dressel2010,Dressel2011} to formally construct system observables from the probability operators \eqref{eq:povm}.  This formalism acknowledges that the only quantities to which we have experimental access are the detector drain probabilities, so all observations we wish to make about the system must be contained somehow in those probabilities.  Generally, the correspondence between the detector drains and a particular system observable will be imperfect, but we can compensate for such \emph{ambiguity} of the detection by weighting the drain probabilities with appropriate values for the particular measurement setup.  

Generally, we cannot construct information about just any system observable from a particular measurement.  To find which observables we \emph{can} measure, it is useful to decompose the probability operators \eqref{eq:povm} into an orthonormal basis for the observable space.  In our case, the system state space is two-dimensional, so any Hermitian operator can be spanned by the four basis operators,
\begin{subequations}\label{eq:basis}
\begin{align}
  \op{\sigma}^s_0 &= \op{1} = \prj{L^s} + \prj{U^s}, \\
  \op{\sigma}^s_1 &= \op{\sigma}^s_x = \oprod{L^s}{U^s} + \oprod{U^s}{L^s}, \\
  \op{\sigma}^s_2 &= \op{\sigma}^s_y = -i\left( \oprod{L^s}{U^s} - \oprod{U^s}{L^s} \right), \\
  \op{\sigma}^s_3 &= \op{\sigma}^s_z = \prj{L^s} - \prj{U^s},
\end{align}
\end{subequations}
which are equivalent to the identity operator and the Pauli spin operators.  To find the real components of an observable in this basis we introduce the normalized Hilbert-Schmidt inner product between operators,
\begin{align}
  \hsp{\op{A}}{\op{B}} = \frac{\Tr{\op{A}^\dagger\op{B}}}{\Tr{\op{1}}},
\end{align}
under which the operator basis is orthonormal, 
\begin{align}
  \hsp{\op{\sigma}^s_\mu}{\op{\sigma}^s_\nu} = \delta_{\mu\nu}.
\end{align}
Here $\mu,\nu\in{0,1,2,3}$ and $\delta_{\mu\nu}$ is the Kronecker delta that is $1$ if $\mu=\nu$ and $0$ otherwise.  Using this basis, any system observable can be written,
\begin{subequations}\label{eq:expansion}
\begin{align}
  \op{A} &= \sum_\mu a_\mu \op{\sigma}^s_\mu, \\
  a_\mu &= \hsp{\op{A}}{\op{\sigma}^s_\mu} = \Tr{\op{A}\op{\sigma}^s_\mu}/2,
\end{align}
\end{subequations}
where $\{a_\mu\}$ are real-valued components of the observable.

Using \eqref{eq:expansion}, we can expand the probability operators on the system \eqref{eq:povm} in the basis \eqref{eq:basis} to determine their structure,
\begin{subequations}\label{eq:povmexpand}
  \begin{align}
    \op{E}_{D_1} &= \frac{1}{2}\left( \beta^d_+ - V^d\Delta^d \right)\op{\sigma}^s_0 - \frac{1}{2}V^d\Gamma^d \,\op{\sigma}^s_3, \\
    \op{E}_{D_2} &= \frac{1}{2}\left( \beta^d_- + V^d\Delta^d \right)\op{\sigma}^s_0 + \frac{1}{2}V^d\Gamma^d \,\op{\sigma}^s_3,
  \end{align}
\end{subequations}
where we see that the measurement is characterized by the detector parameters,
\begin{subequations} \label{eq:params}
\begin{align}
  \beta^d_+    &= 2\left(T^d_1\, T^d_2\,  + R^d_1\, R^d_2\, \right) = 1 + \delta^d_1\, \delta^d_2\, , \displaybreak[0] \\
  \beta^d_-    &= 2\left(T^d_1\, R^d_2\,  + R^d_1\, T^d_2\, \right) = 1 - \delta^d_1\, \delta^d_2\, , \displaybreak[0] \\
  V^d          &= 4\sqrt{T^d_1\, R^d_1\, T^d_2\, R^d_2\, } = \epsilon^d_1\, \epsilon^d_2\, , \displaybreak[0] \\
  \Gamma^d     &= \sin\frac{\gamma}{2}\sin\left(\frac{\gamma}{2}+\phi^d\right), \displaybreak[0] \\
  \Delta^d     &= \cos\phi^d - \Gamma^d. 
\end{align}
\end{subequations}
defined in terms of the QPC balance parameters \eqref{eq:qpcbalance}, the tuning phases \eqref{eq:tuning}, and the coupling phase $\gamma$.  We will describe these parameters in detail in the next section.

The probability operators only contain components in the subspace spanned by $\{\op{\sigma}^s_0,\op{\sigma}^s_3\}$; therefore, \emph{we can only construct observables that are contained within that subspace}.  That is, we can construct any observable of the form $\op{A} = a_0 \op{\sigma}^s_0 + a_3 \op{\sigma}^s_3$.  We denote observables of this form as being \emph{compatible} with the measurement \eqref{eq:povmexpand}.  Other observables are \emph{incompatible} with the measurement.

To construct such a compatible system observable from the measurement, we expand its operator directly in terms of the probability operators \eqref{eq:povm},
\begin{align}
  \label{eq:compat}
  \op{A} &= a_0 \op{\sigma}^s_0 + a_3 \op{\sigma}^s_3 = \alpha_{D_1} \op{E}_{D_1} + \alpha_{D_2} \op{E}_{D_2}.
\end{align}
The required expansion coefficients $\alpha_{D_1}$ and $\alpha_{D_2}$ are generalized eigenvalues, or \emph{contextual values}~\cite{Dressel2010,Dressel2011}, of the operator.  Using this expansion, we can recover the same information \emph{on average} as a projective measurement by using only the drain probabilities, $\mean{\op{A}} = \alpha_{D_1} P_{D_1} + \alpha_{D_2} P_{D_2}$.

To determine the appropriate contextual values to assign in order to construct $\op{A}$, we insert \eqref{eq:povmexpand} into \eqref{eq:compat} and solve it as a standard matrix equation using the orthonormal basis, which yields the unique contextual values,
\begin{subequations}\label{eq:cvalues}
\begin{align}
  \label{eq:cvalued1}
  \alpha_{D_1} &= a_0 - \frac{a_3}{\Gamma^d}\left(\frac{\beta^d_-}{V^d} + \Delta^d \right), \\
  \alpha_{D_2} &= a_0 + \frac{a_3}{\Gamma^d}\left(\frac{\beta^d_+}{V^d} - \Delta^d \right). 
\end{align}
\end{subequations}

As long as the contextual values do not diverge, the expansion \eqref{eq:compat} of the compatible operator $\op{A}$ is well defined, and we can perfectly recover its average,
\begin{align}
  \mean{\op{A}} &= \bra{\psi^s}\op{A}\ket{\psi^s} = a_0 + a_3 \delta^s_1\, .
\end{align}
The observable parameter $a_0$ sets the reference point for the average, so contributes no information about the system; we will set it to zero in what follows without loss of generality.  Similarly, the remaining parameter $a_3$ sets the scale of the average; we will set it to one in what follows.  

We formally conclude that the detector drains perform a generalized measurement of the \emph{which-path operator} $\op{\sigma}^s_3$, as might be intuitively expected from the path-dependent interaction.  Moreover, the QPC parameter $\delta^s_1\, $ defined in \eqref{eq:delta} determines the particle-like which-path behavior on average.  No other information about the system can be inferred from the measurement.

\begin{figure}[t]
  \begin{center}
    \includegraphics[width=8cm]{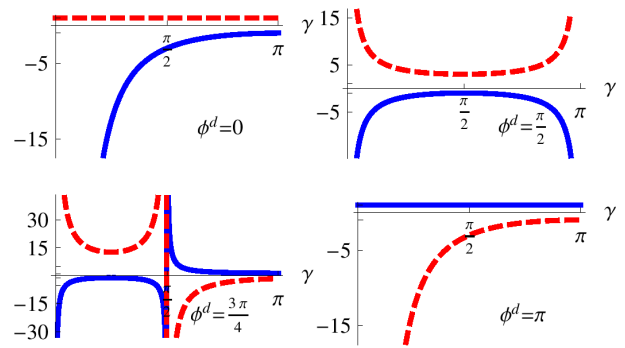}
  \end{center}
  \caption{(color online) The contextual values \eqref{eq:cvalues} of the which-path operator $\op{\sigma}^s_3$: $\alpha_{D_1}$ (solid, blue) and $\alpha_{D_2}$ (dashed, red), as a function of the coupling phase $\gamma$.  The curves are shown for efficient detection $V^d = 1$ and detector tunings $\phi^d = \{0,\pi/2,3\pi/4,\pi\}$.  The tuning strongly affects the ambiguity of the measurement; moreover, the roles of the detector drains flip as the tuning varies from $\phi^d=0$ to $\phi^d=\pi$.}
  \label{fig:cvalues}
\end{figure}

The contextual values \eqref{eq:cvalues} are shown in Figure~\ref{fig:cvalues} for a few parameter choices.  If they are equal to the eigenvalues of the which-path operator, $\alpha_{D_1},\alpha_{D_2} = \pm1$, then the measurement is \emph{unambiguous}: one obtains perfect knowledge about the path information with every drain detection, and the system state is projected to a pure path state.  If the contextual values diverge, $\alpha_{D_1},\alpha_{D_2} \to \pm\infty$, then the measurement is \emph{completely ambiguous}: no knowledge about the path information can be obtained, and the system state is unprojected; however, will shall see in Section~\ref{sec:disturbance} that the system state may still be unitarily perturbed by the coupling.  In between these extremes the measurement is \emph{partially ambiguous}: partial knowledge is obtained about the path information with each drain detection, and the system state is partially projected toward a particular path state.

\subsection{Parameters}
\begin{figure}[t]
  \begin{center}
    \includegraphics[width=8cm]{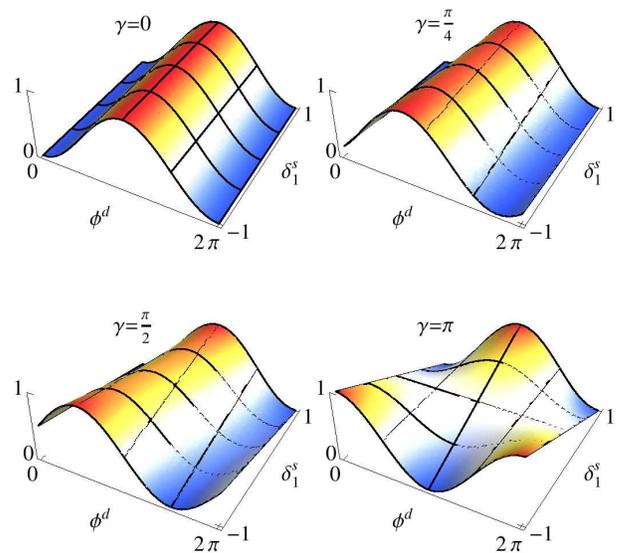}
  \end{center}
  \caption{(color online) The drain probability $P_{D_1}$ \eqref{eq:probsdet} as a function of the detector tuning $\phi^d$ and the which-path information $\delta^s_1$, shown for efficient detection $V^d = 1$ and coupling phases $\gamma=\{0,\pi/4,\pi/2,\pi\}$.  For zero coupling the interference is independent of the which-path information; for strong coupling $\gamma=\pi$ the interference maximally corresponds to the which-path information.}
  \label{fig:drainprobs}
\end{figure}

To better understand the parameters \eqref{eq:params} we write the drain probabilities \eqref{eq:probs} explicitly,
\begin{subequations} \label{eq:probsdet}
\begin{align}
  P_{D_1} &= \frac{1}{2} \left(\beta^d_+ - V^d \left( \Delta^d + \delta^s_1\,  \Gamma^d \right)\right), \\
  P_{D_2} &= \frac{1}{2} \left(\beta^d_- + V^d \left( \Delta^d + \delta^s_1\,  \Gamma^d \right)\right).
\end{align}
\end{subequations}
The probability $P_{D_1}$ is illustrated in Figure~\ref{fig:drainprobs} for several values of the coupling strength.

The particle-like parameters $\beta^d_+,\beta^d_-\in[0,1]$ are determined entirely by the path-bias parameters $\delta^d_1\, $ and $\delta^d_2\, $; they indicate the average background signal of each detector drain and satisfy $(\beta^d_++\beta^d_-)/2 = 1$.  The wave-like parameter $V^d\in[0,1]$ is determined entirely by the path-uncertainty parameters $\epsilon^d_1\, $ and $\epsilon^d_2\, $; it indicates the \emph{visibility} of the interference.  The parameter $\Gamma^d\in[-1,1]$ indicates the deviation in the interference caused by the coupling phase $\gamma$, which is the only effect of the charge coupling.  The parameter $\Delta^d\in[-1,1]$ indicates the interference unrelated to the path information of the system.  As the coupling $\gamma\to 0$, then $\Gamma^d\to 0$ and $\Delta^d\to\cos\phi^d$, which recovers the signal for an isolated interferometer.~\cite{Ji2003}  As the coupling $\gamma\to \pi$, then $\Gamma^d\to \cos\phi^d$ and $\Delta^d\to 0$, and the interference maximally corresponds to the path information.

The parameters \eqref{eq:params} also give insight into the nature of the \emph{measurement} by the role they play in the \emph{contextual values} \eqref{eq:cvalues}.  The parameter $\Gamma^d$ indicates the correlation between the detector drains and the which-path information.  Its magnitude $|\Gamma^d|\in[0,1]$ denotes the \emph{correlation strength}, with $1$ indicating perfect correlation and $0$ indicating no correlation; due to the inverse dependence in \eqref{eq:cvalues}, any imperfect correlation will \emph{amplify} the contextual values to compensate for the resulting measurement ambiguity.  The sign of $\Gamma^d$ indicates the correspondence of the detector drains to the which path information, with $-$ denoting the mapping $\{D_1,D_2\} \leftrightarrow \{L^s,U^s\}$ and $+$ denoting the mapping $\{D_2,D_1\} \leftrightarrow \{L^s,U^s\}$.  Note that the correlation strength depends not only on the coupling phase $\gamma$, but also on the tuning phase $\phi^d$; hence, it is possible for the detector drains to be uncorrelated with the system paths even under strong coupling (e.g. examine $\phi^d = \pi/2$ in Figure~\ref{fig:drainprobs} when $\gamma=\pi$). 

The parameters $\beta^d_+$ and $\beta^d_-$ in \eqref{eq:cvalues} counterbalance the bias in the average drain background caused by a preferred particle-like path.  For instance, if $\beta^d_+ > \beta^d_-$ then the signal at drain $D_1$ is stronger on average in \eqref{eq:probsdet}; hence, the contextual value \eqref{eq:cvalued1} assigned to $D_1$ is proportional to the smaller value $\beta^d_-$ to compensate.

The visibility parameter $V^d$ controls the wave-like interference produced by $Q^d_1$ and $Q^d_2$.  The transmission of each QPC should be balanced in order to provide the interaction phase with an equal-amplitude reference phase for later interference.  Any imbalance leads to \emph{inefficiency} of the measurement\cite{Buttiker2002} by reducing the interference visibility, effectively hiding the correlations.  Such inefficiency increases the measurement \emph{ambiguity} and results in an amplification of the contextual values.  All correlations are hidden at zero interference visibility when either $Q^d_1$ or $Q^d_2$ is fully transmissive or reflective, $T^d_1\, ,T^d_2\, \in\{0,1\}$, which leads to divergent contextual values.  Maximum interference visibility occurs for balanced transmission with $V^d = 1$.  We see that to optimally measure the particle-like which-path information for the system, the detector must itself exhibit maximal wave-like interference; \emph{the detector and system behaviors are therefore complementary}.  

The parameter $\Delta^d$ is the portion of the interference not affected by the coupling, meaning  $\Delta^d + \Gamma^d = \cos\phi^d$. It indicates an additional bias in the drain correspondence caused by the interference not pertinent to the which-path measurement.  The contextual values naturally subtract the contribution from this irrelevant background interference to retrieve the measurement information.  In the limit of strong coupling $\gamma\to\pi$, \emph{all} the detector interference encodes the measurement result: $\Gamma^d\to\cos\phi^d$ and $\Delta^d\to0$.

Practically speaking, the detector must be \emph{calibrated} in the laboratory before it can be used to probe an unknown system state.  That is, the detector parameters \eqref{eq:params} must be predetermined by examining the drain outputs of the detector under known system configurations.  For example, pinching off QPC $Q^s_1$ to prevent any interactions allows most of the parameters to be set directly by tuning the detector QPCs and the magnetic field.  The remaining interaction parameter $\gamma$ can be inferred from an additional reference system state.  Therefore, the process of detector calibration can be viewed as the experimental determination of the appropriate \emph{contextual values} to assign to the detection apparatus.

\subsection{Measurement Disturbance}\label{sec:disturbance}
The measurement necessarily disturbs the system state by extracting information.  We can see the effect of such disturbance by characterizing the system interferometer with analogous parameters to \eqref{eq:params},
\begin{subequations}\label{eq:paramssys}
\begin{align}
  \beta^s_+    &= 2\left(T^s_1\, T^s_2\,  + R^s_1\, R^s_2\, \right) = 1 + \delta^s_1\, \delta^s_2\, , \displaybreak[0] \\
  \beta^s_-    &= 2\left(T^s_1\, R^s_2\,  + R^s_1\, T^s_2\, \right) = 1 - \delta^s_1\, \delta^s_2\, , \displaybreak[0] \\
  V^s          &= 4\sqrt{T^s_1\, R^s_1\, T^s_2\, R^s_2\, } = \epsilon^s_1\, \epsilon^s_2\, , \displaybreak[0] \\
  \Gamma^s     &= \sin\frac{\gamma}{2}\sin\left(\frac{\gamma}{2} - \phi^s\right), \displaybreak[0] \\
  \Delta^s     &= \cos\phi^s - \Gamma^s.
\end{align}
\end{subequations}
Using these parameters the absorption probabilities for the system drain take the simple form similar to \eqref{eq:probsdet},
\begin{subequations}\label{eq:probssys}
  \begin{align}
    P_{S_1} &= \frac{1}{2}\left( \beta^s_+ - V^s\left(\Delta^s - \delta^d_1\,  \Gamma^s \right) \right), \\
    P_{S_2} &= \frac{1}{2}\left( \beta^s_- + V^s\left(\Delta^s - \delta^d_1\,  \Gamma^s \right) \right).
  \end{align}
\end{subequations}

With efficient detection $V^d=1$ and strong coupling $\gamma\to\pi$, then $\delta^d_1 \to 0$ and $\Delta^s\to 0$, so the system drain probabilities display no interference, $P_{S_1}\to\beta^s_+/2$ and $P_{S_2}\to\beta^s_-/2$; that is, a strongly coupled, efficient which-path measurement will force particle-like statistics in the system.~\cite{Neder2007a,Kang2007}  

The measurement disturbance may be analyzed more explicitly by rewriting the measurement operators \eqref{eq:measops} for the case of efficient detection $V^d = 1$,
\begin{subequations}\label{eq:effmeasops}
\begin{align}
  \op{M}_{D_1} &= i\, e^{i\chi^d_2}\, e^{i\phi^d/2} \, \op{U}_\gamma \, \op{E}_{D_1}^{1/2}, \\
  \op{M}_{D_2} &= i\, e^{i\xi^d_2}\, e^{i\phi^d/2} \, \op{U}_\gamma \, \op{E}_{D_2}^{1/2}, \\
  \op{E}_{D_1}^{1/2} &= \sin\frac{\phi^d}{2}\, \prj{L^s} + \sin\frac{\phi^d + \gamma}{2}\, \prj{U^s}, \\
  \op{E}_{D_2}^{1/2} &= \cos\frac{\phi^d}{2}\, \prj{L^s} + \cos\frac{\phi^d + \gamma}{2}\, \prj{U^s}, \\
  \label{eq:udisturb}
  \op{U}_\gamma &= \exp\left(i \frac{\gamma}{2}\prj{U^s}\right).
 \end{align}
\end{subequations}
The disturbance manifests itself as two distinct processes.  First, the positive roots of the POVM \eqref{eq:povm} $\{\op{E}_{D_1}^{1/2},\op{E}_{D_2}^{1/2}\}$ perform the information extraction necessary for the measurement, partially projecting the reduced system state toward a particular path.  Second, the coupling-dependent unitary factor $\op{U}_\gamma$ contributes an additional evolution of the system that is unrelated to the extraction of information.  The remaining phase factors contribute only to the global phase of the measured state and do not alter the subsequent measurement statistics.

Unambiguous measurements extract maximal information from the system and thus project the system state to a definite path; they are frequently known as \emph{projective} or \emph{strong} measurements.  Ambiguous measurements extract partial information from the system and thus partially project the system state toward a particular path.  Completely ambiguous measurements extract no information from the system and thus are completely unitary.  When the system state is nearly unperturbed up to a global phase, the measurement is called \emph{weak}, which corresponds to the case of a nearly completely ambiguous measurement with a negligible unitary evolution.

\subsection{Strong Coupling}\label{sec:strong}
An unambiguous measurement can only be obtained in the limits of efficient detection $V^d \to 1$ and strong coupling $\gamma\to\pi$.  In this situation, the ambiguity will be determined only by the tuning phase of the detector $\phi^d$, and the POVM will have the most symmetric dependence on the which-path operator,
\begin{subequations}
\begin{align}
  \alpha_{D_1} &\to \frac{-1}{\cos\phi^d}, \\
  \alpha_{D_2} &\to \frac{1}{\cos\phi^d},  \\
  \op{E}_{D_1} &\to \frac{1}{2}\left( \op{1} - \op{\sigma}^s_3 \cos\phi^d \right), \\
  \op{E}_{D_2} &\to \frac{1}{2}\left( \op{1} + \op{\sigma}^s_3 \cos\phi^d \right).
\end{align}
\end{subequations}

As the tuning phase $\phi^d$ varies, the POVM elements oscillate between pure path projections and the identity, despite the strong coupling.  The tuning-dependent drain ambiguity contributes to the inefficiency of the measurement by erasing the potentially extractable which-path information from the detector state.  Indeed, we shall see in Section~\ref{sec:erasure} that such ambiguity in the measurement allows the system interference to be recovered by conditioning the system results on specific detector outcomes:  Such a phenomenon is known as \emph{quantum erasure}.\cite{Wheeler1979,Scully1982,Scully1991,Herzog1995,Mir2007,Hilmer2007} 

In a laboratory quantum Hall system the AB phase will precess due to slow decay of the transverse magnetic field, so the tuning phase $\phi^d$ will also precess slowly.  Hence, the ambiguity of the measurement will generally oscillate between extremes, while also flipping the correspondence of the drains to the which-path information.  Despite any ambiguity in the measurement, however, the system will always be perturbed by the additional unitary evolution \eqref{eq:udisturb} that induces a relative phase shift of $\pi/2$ between the arms.  Since the system state will be appreciably altered by the strong coupling, the measurement will not be weak even when completely ambiguous.  Hence, \emph{ambiguity of the measurement need not indicate weakness of the measurement}.

The measurement becomes \emph{unambiguous} when the tuning is held fixed at $\cos\phi^d=\pm1$.  In this situation, the detector drains are perfect ``bright'' and ``dark'' ports: detection at the dark port will occur deterministically when the system excitation is in the upper arm.  The detector drains are perfectly correlated to the which-path information of the system, so the system state is projected to a definite path, and the measurement is \emph{strong}.

\subsection{Weak Coupling Limit}\label{sec:weak}
The weak coupling limit is the limit as the coupling phase $\gamma \to 0$ and the system and detector become nearly uncoupled.  Since at zero coupling the measurement operators \eqref{eq:effmeasops} must either be zero or be proportional to the identity, the weak coupling limit of a measurement must have outcomes that are inherently ambiguous.  Hence, we expect the contextual values \eqref{eq:cvalues} to diverge.  However, since $\Gamma^d = \sin\phi^d (\gamma/2) + \cos\phi^d (\gamma/2)^2 + O(\gamma^3)$ the nature of the divergence will also depend upon the tuning.  

If the tuning is not an integer multiple of $\pi$, then both measurement operators \eqref{eq:effmeasops} will approach the identity as $\gamma\to 0$ and the measurement will be \emph{weak} for all outcomes.  That is, the system state will be nearly unperturbed for any outcome of the measurement.  In this case, $\Gamma^d = \sin\phi^d (\gamma/2) + O(\gamma^2)$ and the divergence of the contextual values \eqref{eq:cvalues} will be linear in $\gamma$.  For an efficient detector with $V^d = 1$, we find to $O(\gamma^2)$, 
\begin{subequations}\label{eq:weak}
  \begin{align}
    \alpha_{D_1} &\to 1 - \frac{2}{\gamma}\frac{1 + \cos\phi^d}{\sin\phi^d}, \\
    \alpha_{D_2} &\to 1 + \frac{2}{\gamma}\frac{1 - \cos\phi^d}{\sin\phi^d}, \\
    \op{E}_{D_1} &\to \frac{1 - \cos\phi^d}{2}\op{1} + \frac{\gamma}{2}\sin\phi^d \prj{U^s}, \\
    \op{E}_{D_2} &\to \frac{1 + \cos\phi^d}{2}\op{1} - \frac{\gamma}{2}\sin\phi^d \prj{U^s}.
  \end{align}
\end{subequations}
The POVM has simple dependence on the projection to the upper path, which can also be written in terms of the which-path operator, $\prj{U^s} = (\op{1} - \op{\sigma}^s_3)/2$.  The most symmetric case of $\phi^d = \pi/2$ is shown in the upper-right of Figure~\ref{fig:cvalues}.

However, if the tuning $\phi^d = n\pi$ with integer $n$, then \emph{only one} of the measurement operators will approach the identity as $\gamma\to 0$.  The remaining outcome remains proportional to a projector with a vanishing coefficient and will thus strongly perturb the system state.  Hence, only one contextual value diverges while the other remains a constant eigenvalue.  In this case, $\Gamma^d = (-1)^n \sin^2(\gamma/2)$ so the divergence will be quadratic in $\gamma$.  We call such a measurement a \emph{semi-weak} measurement \cite{Dressel2011} since only a subset of outcomes are weak.  For an efficient detector with $V^d = 1$, we find,
\begin{subequations}\label{eq:semiweak}
  \begin{align}
    \alpha_{D_1} &\to \frac{-1}{\sin^2\frac{\gamma}{2}}\left((-1)^n + \cos^2\frac{\gamma}{2}\right), \\
    \alpha_{D_2} &\to \frac{1}{\sin^2\frac{\gamma}{2}}\left((-1)^n - \cos^2\frac{\gamma}{2}\right),  \\
    \op{E}_{D_1} &\to \frac{1}{2}(1 - (-1)^n)\op{1} + (-1)^n \sin^2\frac{\gamma}{2} \prj{U^s}, \\
    \op{E}_{D_2} &\to \frac{1}{2}(1 + (-1)^n)\op{1} - (-1)^n \sin^2\frac{\gamma}{2} \prj{U^s}.
  \end{align}
\end{subequations}
The POVM retains the simple dependence on the projection to the upper path.  The cases for $n=0$ and $n=1$ are shown in the upper-left and lower-right of Figure~\ref{fig:cvalues}, respectively.

For the semi-weak measurement, the effect of absorption at one of the drains is projective.  The projective drain outcome unambiguously indicates that the system excitation took the upper $U^s$ path; therefore, the contextual value assigned to the \emph{complementary} drain is an eigenvalue.  In contrast, the effect of absorption at the complementary drain only weakly perturbs the system state.  Its outcome only ambiguously corresponds to which-path information; therefore, the contextual value assigned to the \emph{projective} drain must be amplified.  Such complementary behavior of the contextual value amplification can be counter-intuitive, but it emphasizes that the function of the amplification is to compensate for the ambiguity of the measurement.

We shall see in Section~\ref{sec:weakvalue} that while conditioned averages of the weak measurements \eqref{eq:weak} will lead to \emph{weak values}, the conditioned averages of the semi-weak measurements \eqref{eq:semiweak} have different limiting behavior and lead to different values.  The two limiting cases will compete depending on the relative magnitudes of $\gamma$ and $\phi^d$.

\subsection{Fluctuating Coupling}\label{sec:fluctuation}
\begin{figure}[t]
  \begin{center}
    \includegraphics[width=\columnwidth]{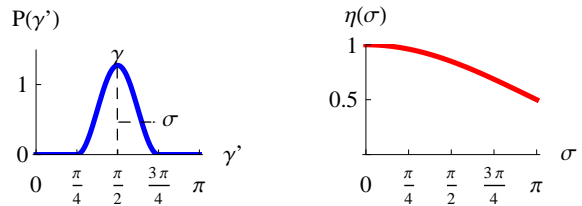}
  \end{center}
  \caption{(color online) Left: The raised cosine distribution showing a spread in the coupling phase centered at $\gamma=\pi/2$ by a half-width $\sigma=\pi/4$.  Right: The inefficiency factor $\eta(\sigma)$ defined in \eqref{eq:eta} as a function of the half-width $\sigma$.}
  \label{fig:cosine}
\end{figure}

If the coupling between excitation pairs is not a constant relative phase $\gamma$, but instead can fluctuate within a finite uncertainty width $\sigma$ around an average $\gamma$, then the average measurement will be correspondingly more ambiguous.  We could quantify this effect by averaging the joint state \eqref{eq:scatterstate} over a range of coupling phases to create a mixed state represented by a density operator; the measurement operators \eqref{eq:measops} and resulting POVM \eqref{eq:povm} could then be generalized to an \emph{averaged} measurement from that density operator.  However, that procedure would be completely equivalent to the simpler procedure of averaging the probability operators \eqref{eq:povm} over the coupling width directly, which we choose to do here.

For simplicity, we consider as a coupling distribution the \emph{raised cosine distribution}, which is Gaussian-like, but has compact support.  We center the distribution around $\gamma\in[0,2\pi]$, and give it the half-width $\sigma\in[0,\pi]$.  The density for the distribution is nonzero in the domain $\gamma'\in[\gamma-\sigma,\gamma+\sigma]$ and has the form,
\begin{align}
  P(\gamma') &= \frac{1}{2\sigma}\left(1 + \cos\left(\frac{\pi}{\sigma}(\gamma' - \gamma) \right)\right). 
\end{align}
An example of the distribution is shown in Figure~\ref{fig:cosine} centered at $\gamma=\pi/2$ and with half-width $\sigma=\pi/4$.

Averaging the probability operators \eqref{eq:povm} only affects the constant $\Gamma^d$, which is replaced by the averaged version,
\begin{align}
  \Gamma^d(\gamma) &\to \int_{\gamma-\sigma}^{\gamma+\sigma}d\gamma'\, P(\gamma') \Gamma^d(\gamma') = \eta(\sigma) \Gamma^d(\gamma), \\
  \label{eq:eta}
  \eta(\sigma) &= \left( \frac{\pi^2}{\pi^2 - \sigma^2} \right)\left( \frac{\sin\sigma}{\sigma} \right).
\end{align}
A plot of the damping factor $\eta(\sigma)$ is shown in Figure~\ref{fig:cosine}.  

We assumed in \eqref{eq:initstate} that the source emits only excitation pairs.  However, any contribution of \emph{unpaired} excitations in the initial joint state is equivalent to a contribution of joint states with $\gamma=0$.  The net effect of the source emitting such unpaired excitations is thus to modify $\Gamma^d$ by an additional probability factor $P_p$ that denotes the likelihood of pair emission.  Hence, the only effect of an imperfect source is to introduce a net inefficiency factor $\eta' = P_p\, \eta(\sigma)\in[0,1]$ in the parameter $\Gamma^d$.

Since the contextual values \eqref{eq:cvalues} inversely depend on $\Gamma^d$, any such inefficiency will introduce an overall amplification factor of $1/\eta'$.  In other words, any uncertainty in the coupling strength will lead to additional ambiguity in the average measurement by degrading the portion of the detector interference that is coupled to the system.

\subsection{Observation Time}\label{sec:meastime}
Since ambiguous measurements provide less information about an observable per measurement, more measurements will be required to achieve a desired precision for an observable average.  We can characterize the necessary increase in observation time as follows.  The total detector current is $I = (e^2 V P)/h = e/\tau_m$ according to \eqref{eq:current_approx}, where $P$ is the total probability for excitations to traverse the sample; hence, we can infer that the average time per detector absorption is $\tau_m = h / eV P$.  For our single-particle model to apply we wish for the voltage bias $V$ to be low enough that the interferometers contain less than one excitation per channel on average.  The characteristic measurement time $\tau_m$ will then be on the order of the time-of-flight $\tau_m \approx \ell/v_F$ of an excitation pair through the sample, where $\ell$ is the average path length of the interferometers and $v_F$ is the Fermi velocity of the ballistic excitations.  An observation time of $T$ at the drains $D_1$ and $D_2$ therefore roughly corresponds to $n \approx T/\tau_m$ individual measurement events.  

The contextual values can be used to provide an upper bound for the number of measurement events for a desired root-mean-square (RMS) error in the estimation of the average.  Specifically, to estimate the average $\mean{\op{\sigma}^s_3}$ from a sequence of $n$ random drain absorptions $(d_1, d_2, \cdots, d_n)$, where $d_i \in \{D_1, D_2\}$, one can use an \emph{unbiased estimator} for the average,
\begin{align}\label{eq:unbiasedestimator}
  \text{E}[\op{\sigma}_z] &= \frac{1}{n}\sum_i^n \alpha_{d_i},
\end{align}
that is defined in terms of the \emph{contextual values} assigned to each measurement realization.  As $n\to\infty$, the estimator \eqref{eq:unbiasedestimator} converges to $\mean{\op{\sigma}^s_3} = \alpha_{D_1}P_{D_1} + \alpha_{D_2}P_{D_2}$.  The \emph{mean squared error} (MSE) of this estimator is given by the variance of the contextual values over the number of measurements,
\begin{align}
  \text{MSE}[\text{E}[\op{\sigma}_z]] &= \frac{\alpha_{D_1}^2 P_{D_1} + \alpha_{D_2}^2 P_{D_2} - \mean{\op{\sigma}^s_3}^2}{n}.
\end{align}
Hence, the RMS error $\sqrt{\text{MSE}[\text{E}]}$ scales as $1/\sqrt{n}$ and improves with an increasing number of measurements.

Without prior knowledge of the state, a reasonable upper bound one can make for the MSE is the norm-squared of the contextual values over the number of measurements,
\begin{align}
  \text{MSE}[\text{E}[\op{\sigma}_z]] &\leq \frac{\alpha_{D_1}^2 + \alpha_{D_2}^2}{n}.
\end{align} 
It then follows that to guarantee a maximum desired RMS error $\epsilon$ one needs an observation time on the order of,
\begin{align}
  T \approx \tau_m \frac{\alpha_{D_1}^2 + \alpha_{D_2}^2}{\epsilon^2}.
\end{align}
As the measurement becomes more ambiguous the contextual values become more amplified and so lengthen the observation time necessary to achieve the RMS error of $\epsilon$.  For a strong measurement the upper bound on the observation time is $T \approx 2 \tau_m / \epsilon^2$.

\section{Conditioned Measurements}\label{sec:conditioning}
To gain further insight into the which-path information, we can \emph{condition} the measurement on the subsequent absorption of the system excitation at a specific system drain.  To do this we must obtain the joint transmission probabilities for pairs of detector and system drains.  Conditional probabilities can then be defined in terms of the joint and single transmission probabilities.

As pointed out by Kang\cite{Kang2007} these probabilities are experimentally accessible in the low-bias regime through the zero-frequency cross-correlation noise power between a detector drain $D\in\{D_1,D_2\}$ and a system drain $S\in\{S_1,S_2\}$,
\begin{align}
  \label{eq:noisepower}
  S_{D,S} &\approx 2\frac{e^3 V}{h} (P_{S,D}(E_F) - P_S(E_F) P_D(E_F)).
\end{align}
Hence, knowledge of both the average currents \eqref{eq:currents} and the noise power \eqref{eq:noisepower} allows the determination of both the joint and single transmission probabilities.

\subsection{Joint Scattering}
We can determine the joint probabilities directly in the scattering model by rewriting \eqref{eq:scatterstate} in the basis of the system drains using \eqref{eq:qpcs},
\begin{align}
  \begin{pmatrix}\cre{L^s} \\ \cre{U^s}\end{pmatrix} &= \op{U}^s_2 \begin{pmatrix}\cre{S_1} \\ \cre{S_2}\end{pmatrix}
\end{align}
yielding,
\begin{align}
  \label{eq:fullscatterstate}
  \ket{\Psi'''} &= \Big( C_{D_1,S_1}\cre{D_1}\cre{S_1} + C_{D_1,S_2}\cre{D_1}\cre{S_2} \\
              &\; + C_{D_2,S_1}\cre{D_2}\cre{S_1} + C_{D_2,S_2}\cre{D_2}\cre{S_2} \Big)\ket{0}, \nonumber
\end{align}
up to the same global phase as in \eqref{eq:scatterstate}.  

The relevant joint scattering amplitudes are,\begin{widetext}
\begin{subequations}
\begin{align}
  C_{D_1,S_1} &= e^{i (\chi^d_2\, +\chi^s_2\, )} \left[r^d_1\, r^d_2\, r^s_1\, r^s_2\,  + t^d_1\, t^d_2\, r^s_1\, r^s_2\, e^{i (\phi^d + \gamma)} + r^d_1\, r^d_2\, t^s_1\, t^s_2\, e^{i \phi^s} + t^d_1\, t^d_2\, t^s_1\, t^s_2\, e^{i (\phi^d+\phi^s)}\right], \displaybreak[0] \\
  C_{D_1,S_2} &= e^{i (\chi^d_2\, +\xi^s_2\, )} \left[r^d_1\, r^d_2\, r^s_1\, t^s_2\,  + t^d_1\, t^d_2\, r^s_1\, t^s_2\, e^{i (\phi^d + \gamma)} + r^d_1\, r^d_2\, t^s_1\, r^s_2\, e^{i \phi^s} + t^d_1\, t^d_2\, t^s_1\, r^s_2\, e^{i (\phi^d+\phi^s)}\right], \displaybreak[0] \\
  C_{D_2,S_1} &= e^{i (\xi^d_2\, +\chi^s_2\, )} \left[r^d_1\, t^d_2\, r^s_1\, r^s_2\,  + t^d_1\, r^d_2\, r^s_1\, r^s_2\, e^{i (\phi^d + \gamma)} + r^d_1\, t^d_2\, t^s_1\, t^s_2\, e^{i \phi^s} + t^d_1\, r^d_2\, t^s_1\, t^s_2\, e^{i (\phi^d+\phi^s)}\right], \displaybreak[0] \\
  C_{D_2,S_2} &= e^{i (\xi^d_2\, +\xi^s_2\, )} \left[r^d_1\, t^d_2\, r^s_1\, t^s_2\,  + t^d_1\, r^d_2\, r^s_1\, t^s_2\, e^{i (\phi^d + \gamma)} + r^d_1\, t^d_2\, t^s_1\, r^s_2\, e^{i \phi^s} + t^d_1\, r^d_2\, t^s_1\, r^s_2\, e^{i (\phi^d+\phi^s)}\right]. 
\end{align}
\end{subequations}
\end{widetext}

\begin{figure}[t]
  \begin{center}
    \includegraphics[width=8cm]{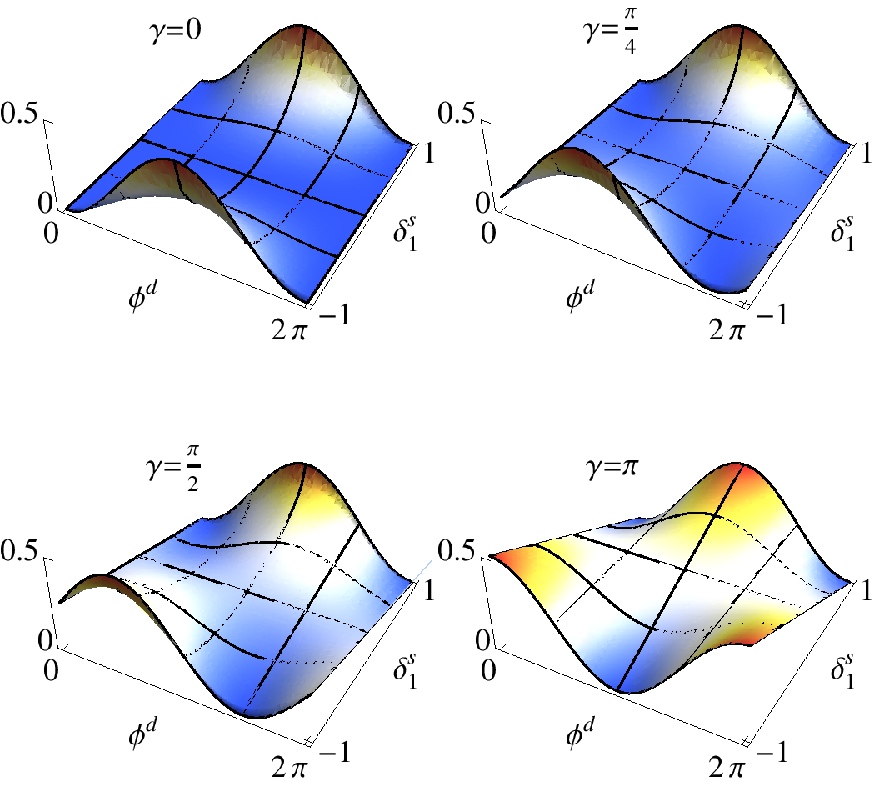}
  \end{center}
  \caption{(color online) The joint probability $P_{D_1,S_1}$ \eqref{eq:jointprobd1s1} with system tuning $\phi^s=0$ as a function of the detector tuning $\phi^d$ and the which-path information $\delta^s_1$, shown for efficient detection $V^d = 1$, balanced system drains $\epsilon^s_2=1$, and the coupling phases $\gamma=\{0,\pi/4,\pi/2,\pi\}$.}
  \label{fig:jointprobs}
\end{figure}

\begin{figure}[t]
  \begin{center}
    \includegraphics[width=8cm]{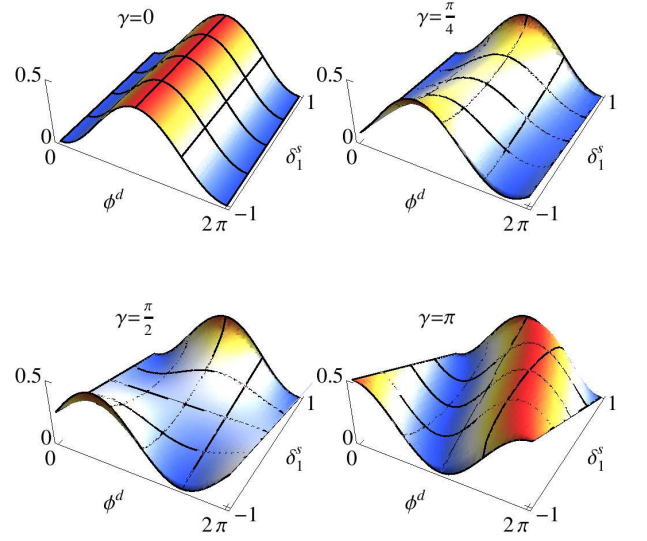}
  \end{center}
  \caption{(color online) The joint probability $P_{D_1,S_1}$ \eqref{eq:jointprobd1s1} with system tuning $\phi^s=\pi/2$ as a function of the detector tuning $\phi^d$ and the which-path information $\delta^s_1$, shown for efficient detection $V^d = 1$, balanced system drains $\epsilon^s_2=1$, and the coupling phases $\gamma=\{0,\pi/4,\pi/2,\pi\}$.}
  \label{fig:jointprobs2}
\end{figure}

The joint probabilities for absorption in detector drain $D\in\{D_1,D_2\}$ and system drain $S\in\{S_1,S_2\}$ can then be understood as either expectations of joint projections $\prj{S}\otimes\prj{D}$ under the joint state \eqref{eq:fullscatterstate}, \emph{or}, equivalently, as expectations of system projections $\prj{S}$ under the \emph{measured} reduced system state $\op{M}_D\ket{\psi^s}$,
\begin{align}\label{eq:jointprob}
  P_{D,S} &= |\iprod{S,D}{\Psi'''}|^2 = |\bra{S}\op{M}_{D}\ket{\psi^s}|^2 = |C_{D,S}|^2.
\end{align}

\begin{figure}[t]
  \begin{center}
    \includegraphics[width=8cm]{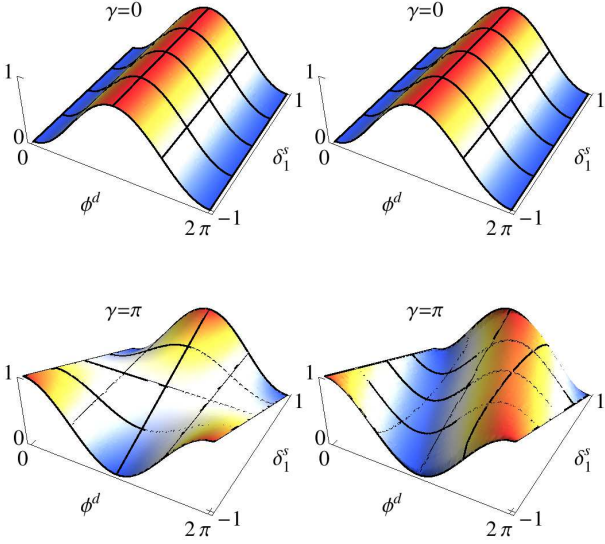}
  \end{center}
  \caption{(color online) The conditional probability $P_{D_1|S_1} = P_{D_1,S_1}/P_{S_1}$ with system tunings $\phi^s=0$ (left) and $\phi^s=\pi/2$ (right) as a function of the detector tuning $\phi^d$ and the which-path information $\delta^s_1$, shown for efficient detection $V^d = 1$, balanced system drains $\epsilon^s_2=1$, and the coupling phases $\gamma=\{0,\pi\}$.}
  \label{fig:condprobs}
\end{figure}

These joint probabilities can be written explicitly in terms of the parameters \eqref{eq:qpcbalance}, \eqref{eq:params}, and \eqref{eq:paramssys} as,
\begin{subequations}\label{eq:jointprobexp}
  \begin{align}\label{eq:jointprobd1s1}
    P_{D_1,S_1} &= \frac{1}{4}\Big( \beta^d_+ \beta^s_+ + V^d V^s \Delta^{ds} \\
    &\quad {} - V^d\left( \Delta^d \beta^s_+ + \Gamma^d\left( \delta^s_1\,  + \delta^s_2\,  \right) \right) \nonumber \\
    &\quad {} - V^s\left( \Delta^s \beta^d_+ - \Gamma^s\left( \delta^d_1\,  + \delta^d_2\,  \right) \right)\Big), \nonumber \displaybreak[0] \\
    P_{D_1,S_2} &= \frac{1}{4}\Big( \beta^d_+ \beta^s_- - V^d V^s \Delta^{ds} \\
    &\quad {} - V^d\left( \Delta^d \beta^s_- + \Gamma^d\left( \delta^s_1\,  - \delta^s_2\,  \right) \right) \nonumber \\
    &\quad {} + V^s\left( \Delta^s \beta^d_+ - \Gamma^s\left( \delta^d_1\,  + \delta^d_2\,  \right) \right)\Big), \nonumber \displaybreak[0] \\
    P_{D_2,S_1} &= \frac{1}{4}\Big( \beta^d_- \beta^s_+ - V^d V^s \Delta^{ds} \\
    &\quad {} + V^d\left( \Delta^d \beta^s_+ + \Gamma^d\left( \delta^s_1\,  + \delta^s_2\,  \right) \right) \nonumber \\
    &\quad {} - V^s\left( \Delta^s \beta^d_- - \Gamma^s\left( \delta^d_1\,  - \delta^d_2\,  \right) \right)\Big), \nonumber \displaybreak[0] \\
    P_{D_2,S_2} &= \frac{1}{4}\Big( \beta^d_- \beta^s_- + V^d V^s \Delta^{ds} \\
    &\quad {} + V^d\left( \Delta^d \beta^s_- + \Gamma^d\left( \delta^s_1\,  - \delta^s_2\,  \right) \right) \nonumber \\
    &\quad {} + V^s\left( \Delta^s \beta^d_- - \Gamma^s\left( \delta^d_1\,  - \delta^d_2\,  \right) \right)\Big), \nonumber 
  \end{align}
\end{subequations}
where we have defined the additional parameter $\Delta^{ds}$ as,
\begin{subequations}
  \begin{align}
    \Delta^{ds} &= \cos\phi^d\cos\phi^s - \Gamma^{ds}, \displaybreak[0] \\
    \Gamma^{ds} &= \sin\frac{\gamma}{2}\sin\left(\frac{\gamma}{2} + \phi^{ds}\right), \displaybreak[0] \\
    \phi^{ds} &= \phi^d - \phi^s.
  \end{align}
\end{subequations}
For illustration purposes, we have plotted the joint probability $P_{D_1,S_1}$ \eqref{eq:jointprobd1s1} for several parameter choices in Figure~\ref{fig:jointprobs} and Figure~\ref{fig:jointprobs2}.

The parameter $\Delta^{ds}\in[-1,1]$ represents the joint interference between the system and detector.  The parameter $\Gamma^{ds}\in[-1,1]$ is the portion of the joint interference that depends explicitly on the coupling phase $\gamma$ and the difference between the tuning phases $\phi^{ds}$.  As the coupling $\gamma\to0$, then $\Gamma^{ds}\to0$ and the joint interference reduces to a decoupled interference product $\Delta^{ds}\to\cos\phi^d\cos\phi^s$.  As the coupling $\gamma\to\pi$, then $\Gamma^{ds}\to\cos\phi^{ds}$ and the joint interference will be maximally coupled $\Delta^{ds}\to-\sin\phi^d\sin\phi^s$.

We can marginalize the joint probabilities \eqref{eq:jointprobexp} to obtain both the detector probabilities \eqref{eq:probsdet} as $P_D = \sum_S P_{D,S}$ and the system probabilities \eqref{eq:probssys} as $P_S = \sum_D P_{D,S}$.  Furthermore, we can construct the \emph{conditional probabilities} $P_{D|S}$ for absorption in a detector drain $D$ given an absorption in a system drain $S$, as well as the conditional probabilities $P_{S|D}$ for absorption in a system drain $S$ given an absorption in a detector drain $D$,
\begin{subequations}\label{eq:condprob}
\begin{align}
  \label{eq:condprobdet}
  P_{D|S} &= \frac{P_{D,S}}{P_{S}}, \\
  \label{eq:condprobsys}
  P_{S|D} &= \frac{P_{D,S}}{P_{D}}.
\end{align}
\end{subequations}
For comparison with the joint probabilities, we have plotted the conditional detector probability $P_{D_1|S_1}$ in Figure~\ref{fig:condprobs} for several parameter choices.

\subsection{Quantum Erasure}\label{sec:erasure}
\begin{figure}[t]
  \begin{center}
    \includegraphics[width=8cm]{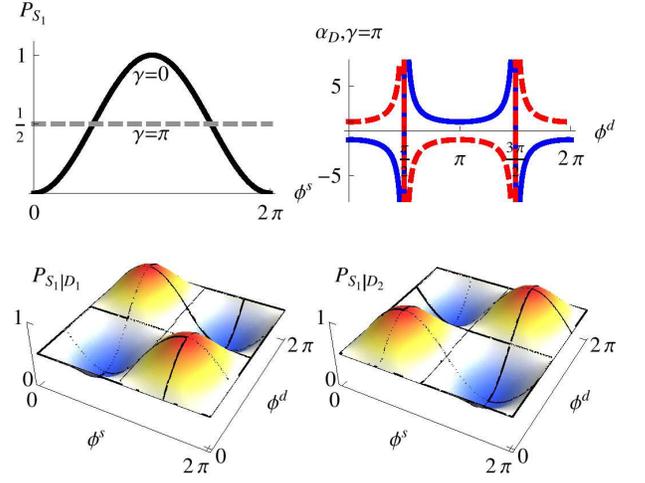}
  \end{center}
  \caption{(color online) Quantum erasure.  Any system interference in the system drain probability $P_{S_1}$ (upper-left, black) is completely destroyed at strong coupling $\gamma=\pi$ (gray, dashed).  Conditioning on the detector drains recovers phase-shifted interference in the conditional probabilities $P_{S_1|D_1}$ and $P_{S_1|D_2}$ (bottom), but with a visibility that is dependent on the measurement ambiguity as controlled by the detector tuning phase $\phi^d$.  The contextual values $\{\alpha_{D_1},\alpha_{D_2}\}$ (upper-right) diverge for maximum ambiguity and reduce to the eigenvalues of the which-path operator $\op{\sigma}^s_3$ for zero ambiguity.  The plots are shown for efficient detection $V^d=1$, strong coupling $\gamma=\pi$, and maximum system visibility $V^s=1$.}
  \label{fig:qeraser}
\end{figure}

We can use the conditional system probabilities $P_{S|D}$ to clarify the phenomenon of \emph{quantum erasure},\cite{Wheeler1979,Scully1982,Scully1991,Herzog1995,Mir2007,Hilmer2007} which has also been explored in this system by Kang\cite{Kang2007}.  Any wave-like interference patterns in the system drain probabilities will degrade with the coupling phase $\gamma$, as shown in the upper-left of Figure~\ref{fig:qeraser} for $P_{S_1}$ and maximum system visibility $V^s = 1$.  At strong coupling $\gamma=\pi$ the wave-like interference will be completely destroyed.  However, parts of the interference may be restored by \emph{conditioning} the drain on an appropriate detector measurement.

To restore the interference in the system statistics, the detector must make an \emph{ambiguous} measurement, as we shall see.  Strong coupling destroys the interference in the unconditioned system statistics by recording the which-path information in the detector state via the coupling phase $\gamma$.  As the which-path information is available in the detector state for later collection, at least in principle, the total reduced system statistics must reflect the degree of \emph{potential} information acquisition.  However, such information in the detector state has not yet been extracted classically since the detector drains have not yet been probed; hence, the information in the state only indicates the \emph{potential} for later extraction of which-path information at the detector drains.  A partially ambiguous measurement extracts some of that potential information and erases the rest; a completely ambiguous measurement extracts no information and erases all of the potential for later extraction in the process.  The recovered interference in the conditioned system statistics reflects the erasure of the information acquisition potential by the ambiguous measurement, even though the total statistics of the reduced system are unchanged by probing the detector.  

As discussed in Section~\ref{sec:strong} the detector phase $\phi^d$ determines the ambiguity of the measurement under such strong coupling, so the degree of possible erasure will also depend on the detector phase.  We can see the dependence of the interference recovery on $\phi^d$ in Figure~\ref{fig:qeraser} in the lower two plots.  As the detector phase $\phi^d$ varies from $0$ to $2\pi$, the conditional probabilities $P_{S_1|D_1}$ and $P_{S_1|D_2}$ continuously vary from flat particle-like statistics to complementary wave-like interference patterns.   The visibilities of the complementary interference patterns directly depend on the measurement ambiguity, as can be seen in the plot of the contextual values in the upper-right of Figure~\ref{fig:qeraser}.  Maximum visibility corresponds to maximum measurement ambiguity where the contextual values diverge; zero visibility corresponds to zero measurement ambiguity where the contextual values reduce to the eigenvalues of the which-path operator.  We also note that the effect of the additional coupling evolution \eqref{eq:udisturb} creates a $\pi/2$ phase shift in the interference pattern that \emph{cannot be erased since it is not part of the information extraction of the measurement}.

\begin{figure}[t]
  \begin{center}
    \includegraphics[width=8cm]{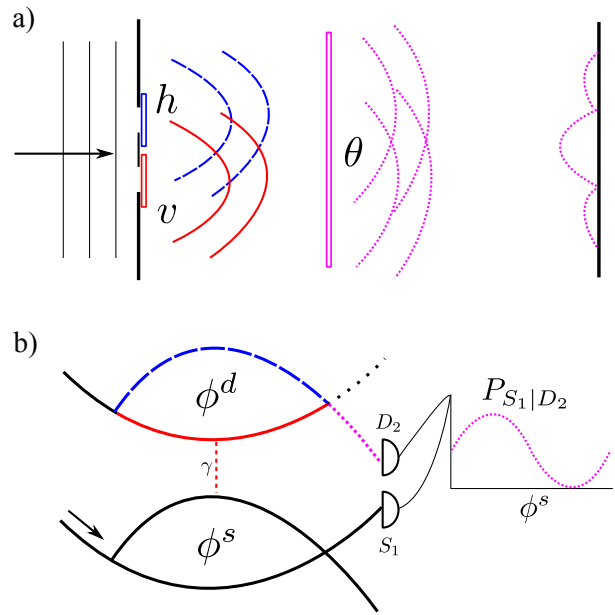}
  \end{center}
  \caption{(color online) Analogy to optical quantum erasure of two-slit interference.  a) A coherent light beam passes through two slits tagged with horizontal (h) and vertical (v) polarization.  After conditioning on a subsequent polarizer oriented at an angle $\theta$ to horizontal, an interference pattern can be recovered with a visibility that depends upon $\theta$.  b) A coherent electron passes through a QPC and is tagged by a detector electron via an interaction phase shift $\gamma$.  After conditioning on the drain $D_2$ after a subsequent detector QPC that forms an MZI with tuning phase $\phi^d$, an interference pattern can be recovered in the system drain $S_1$ as the system tuning $\phi^s$ is varied with a visibility that depends upon $\phi^d$.}
  \label{fig:qeraser2}
\end{figure}

Such erasure under strong coupling has an intuitive analogy to an optical double-slit experiment (that you can even try at home!) \cite{Hilmer2007} as shown in Figure~\ref{fig:qeraser2}.  In the optical equivalent, a coherent beam of light passes through two slits and displays an interference pattern on a remote screen.  However, if an experimenter tags each slit with horizontal and vertical polarizing filters, then the which-path information of the light can later be extracted from the polarization degree of freedom; hence, the total interference pattern on the remote screen will be destroyed.  The experimenter can subsequently condition the measurement on a particular polarization by placing another polarizer after the two slits oriented at some angle $\theta$ relative to horizontal.  If the conditioning polarizer is oriented horizontally or vertically, then the path measurement will be unambiguous and extract all which-path information, so no conditioned system interference will be recovered.  However, if the conditioning polarizer is oriented diagonally, then the path measurement will be completely ambiguous and the potential which-path information will be erased, recovering all of the interference in the conditioned statistics.  In the electronic version, the system excitation plays the role of the light beam, the relative coupling phase $\gamma$ records the potential path information, the conditioning on a particular detector drain plays the role of the polarizer, and the tuning phase $\phi^d$ selects the conditioning basis analogously to the angle $\theta$, controlling the ambiguity of the measurement and degree of erasure.  

Furthermore, one could in principle implement a delayed-choice \cite{Wheeler1979,Scully1982,Herzog1995} version of the quantum erasure by placing the detector drains much further away in the sample than the system drains.  The interaction phase $\gamma$ could be recorded and the system excitations collected, upon which a controlled change in the magnetic field could set the tuning phase of the detector.  Upon conditioning the data, the interference would reappear according to which tuning phase had been chosen after the system excitation had already been collected.

We stress that the erasure of the potential which-path information and recovery of the system interference will be apparent only when conditioning the collected data.  Without conditioning, even completely ambiguous measurements under strong coupling will destroy the system interference.  The interference patterns recovered by conditioning on the detector drains will be complementary to each other in such a case and thus cancel in the unconditioned statistics.

\subsection{Conditioned Averages}
We can also use the conditional probabilities to condition the averages of the which-path measurement on a subsequent \emph{system} drain absorption.  To do this, we weight the conditional detector probabilities \eqref{eq:condprob} with the contextual values for the measurement \eqref{eq:cvalues},~\cite{Dressel2010}
\begin{align}
  \cmean{S}{\op{\sigma}^s_3} &= \sum_D \alpha_{D} P_{D|S}.
\end{align} 
In terms of the various characterization parameters, these take the explicit form,
\begin{subequations}\label{eq:caverages}
  \begin{align}\label{eq:cavs1}
    \cmean{S_1}{\op{\sigma}^s_3} &= \frac{1}{2 P_{S_1}}\left( \delta^s_1\,  + \delta^s_2\,  - V^s\frac{\Xi^{ds}}{\Gamma^d}  \right), \displaybreak[0] \\
    \cmean{S_2}{\op{\sigma}^s_3} &= \frac{1}{2 P_{S_2}}\left( \delta^s_1\,  - \delta^s_2\,  + V^s \frac{\Xi^{ds}}{\Gamma^d} \right),
  \end{align}
\end{subequations}
where there is a joint interference contribution to the conditioned average,
\begin{align}\label{eq:xi}
    \Xi^{ds} &= \Delta^{ds} - (\Delta^s - \delta^d_1\, \Gamma^s) \\
    & + \frac{\delta^d_1\, \delta^d_2\, }{V^d}\left( \Delta^s\beta^d_- - \Gamma^s\left( \delta^d_1\,  + \delta^d_2\,  \right)\right). \nonumber
\end{align}

\begin{figure}[t]
  \begin{center}
    \includegraphics[width=8cm]{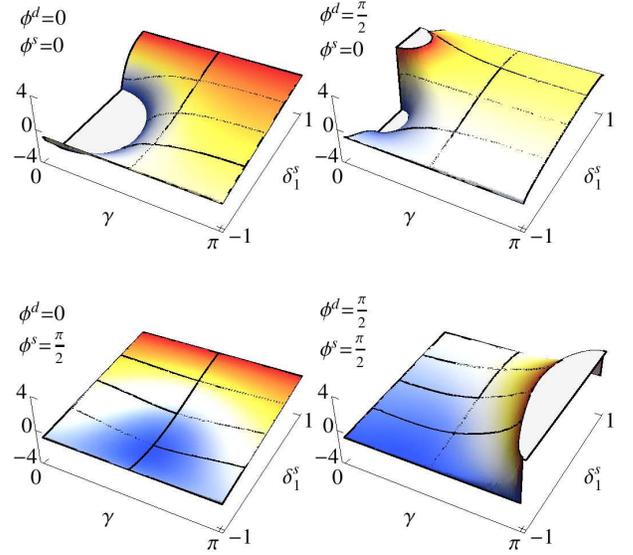}
  \end{center}
  \caption{(color online) The conditioned average $\cmean{S_1}{\op{\sigma}^s_3}$ \eqref{eq:cavs1} as a function of the coupling phase $\gamma$ and the which-path information $\delta^s_1$, shown for efficient detection $V^d = 1$, balanced system drains $\epsilon^s_2=1$, and tuning phases $\phi^d,\phi^s=\{0,\pi/2\}$.}
  \label{fig:caverages}
\end{figure}

The joint interference simplifies considerably in the special case of efficient detection $V^d=1$,
\begin{align}\label{eq:xisimp}
  \frac{\Xi^{ds}}{\Gamma^d} &\to 2\sin\frac{\gamma}{2}\cot\left(\frac{\gamma}{2}+\phi^d\right)\cos\left(\frac{\gamma}{2}-\phi^s\right).
\end{align}
This case is plotted in Figure~\ref{fig:caverages}.

The interference term is scaled by the visibility of the \emph{system} interference $V^s$, which measures the wave-like behavior of the system excitation.  Any wave-like behavior of the system leads to a contribution to the conditioned averages that depends on properties of the correlated \emph{detector} as well, due to the joint interference.  Hence, \emph{the which-path information of the system cannot be decoupled from the detector that is measuring it in general}.

The conditioned averages properly obey the consistency relation $\mean{\op{\sigma}^s_3} = \sum_S \cmean{S}{\op{\sigma}^s_3}\, P_S$, and are bounded by the \emph{contextual values} \eqref{eq:cvalues}.  Since the contextual values are usually larger than the eigenvalues of $\op{\sigma}^s_3$ due to amplification from measurement ambiguity, the conditioned averages can counter-intuitively lie outside the eigenvalue range.  In the weak coupling limit $\gamma\to0$, such conditioned averages can become detector-independent and converge to \emph{weak values},~\cite{Aharonov1988,Dressel2010} as we will show later.

However, for any macroscopic property such conditioned averages will always lie inside the eigenvalue range, even when measured ambiguously.  Specifically, the eigenvalue range constraint for conditioned averages has been shown to be equivalent to a generalized Leggett-Garg constraint~\cite{Leggett1985,Williams2008,Dressel2011} that must be satisfied for any non-invasively measured, realistic property.  As such, any violation of the eigenvalue range in a conditioned average can be seen as a signature of \emph{nonclassical} behavior stemming from quantum interference.

\subsection{Deterministic Measurement}
If either of the system QPCs is fully transmissive or reflective then $V^s=0$, the system interference vanishes, and the conditioned averages \eqref{eq:caverages} reduce to $\pm 1$ for any coupling strength.  In such a case the excitation path is deterministic and the system displays purely particle-like behavior. The post-selection perfectly determines the path, and the which-path measurement made by the detector will always agree with the post-selected value.

\subsection{Strong Coupling}
For the case of strong coupling $\gamma=\pi$ and an efficient detector $V^d=1$, the conditioned averages \eqref{eq:caverages} reduce to,
\begin{subequations}
\begin{align}
  \cmean{S_1}{\op{\sigma}^s_3} &\to \frac{\delta^s_1\,  + \delta^s_2\, }{\beta^s_+} + \frac{V^s}{\beta^s_+}\tan\phi^d\sin\phi^s, \\
  \cmean{S_2}{\op{\sigma}^s_3} &\to \frac{\delta^s_1\,  - \delta^s_2\, }{\beta^s_-} - \frac{V^s}{\beta^s_-}\tan\phi^d\sin\phi^s.
\end{align}
\end{subequations}
The tuning phase $\phi^d$ is the sole detector parameter that specifies the ambiguity of the measurement.  

If the measurement is also unambiguous $\phi^d\to n\pi$, then the interference contribution vanishes.  The conditioned averages become the \emph{detector-independent} quantities $\cmean{S_1}{\op{\sigma}^s_3}\to(\delta^s_1\, +\delta^s_2\, )/\beta^s_+$ and $\cmean{S_2}{\op{\sigma}^s_3}\to(\delta^s_1\, -\delta^s_2\, )/\beta^s_-$ that always lie in the eigenvalue range.  A strong which-path measurement made by the detector therefore forces the system excitation to display particle-like conditioned statistics.

However, even with strong coupling any ambiguity introduced in the detector will lead to quantum erasure that recovers interference in the conditioned statistics of the system, as discussed in Section~\ref{sec:erasure}.  Such recovered interference can take the conditioned averages of the which-path information outside the eigenvalue range.  For an almost completely ambiguous measurement the tuning phase will deviate from $\pi/2$ only by a small angle $\delta\phi^d$.  Since $\tan(\pi/2+\delta\phi^d)=-1/\delta\phi^d+O(\delta \phi^d)$, the interference contribution will dominate, and the conditioned averages will \emph{diverge}.  

\subsection{Weak Coupling Limit}\label{sec:weakvalue}
\begin{figure}[t]
  \begin{center}
    \includegraphics[width=8cm]{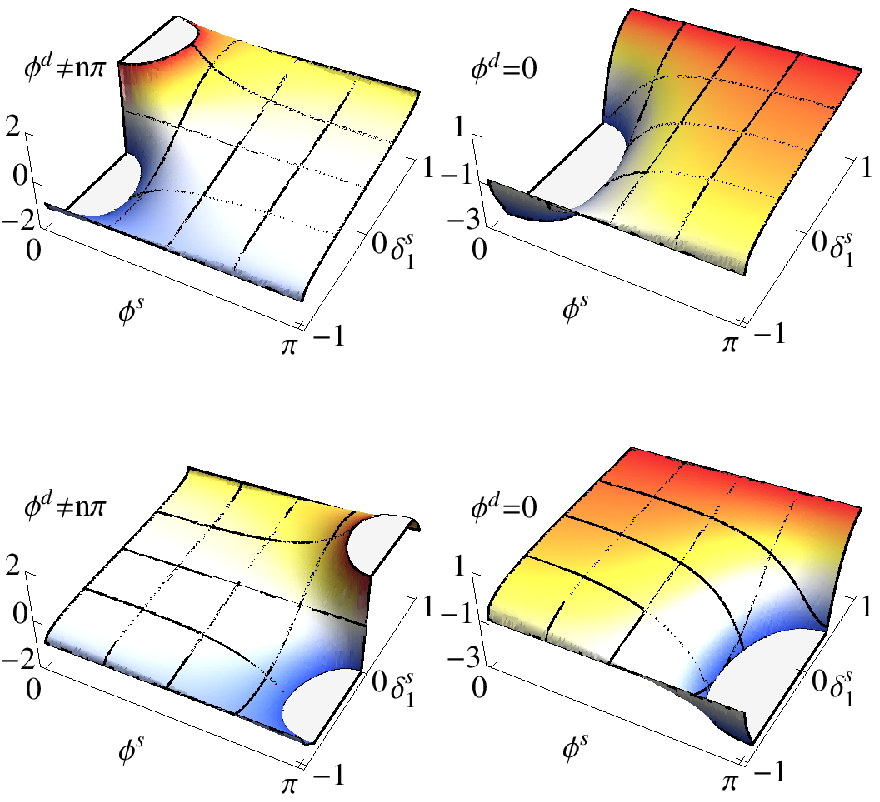}
  \end{center}
  \caption{(color online) The weak limit $\gamma\to0$ of the conditioned averages $\cmean{S_1}{\op{\sigma}^s_3}$ (top) and $\cmean{S_2}{\op{\sigma}^s_3}$ (bottom) for the distinct cases of weak measurement $\phi^d\neq0$ (left) given in \eqref{eq:wvalue} and semi-weak measurement $\phi^d=0$ (right) given in \eqref{eq:semiwvalue} as a function of the system tuning $\phi^s$ and the which-path information $\delta^s_1$. The values are shown for efficient detection $V^d = 1$ and balanced system drains $\epsilon^s_2=1$.}
  \label{fig:wvalues}
\end{figure}

The weak coupling limit $\gamma\to 0$ of an efficient detector $V^d=1$ leads to conditioned averages \eqref{eq:caverages} that generally depend on the detector tuning $\phi^d$, as anticipated during the discussion in Section~\ref{sec:weak}.  For the weak measurement case when the detector tuning is not an integer multiple of $\pi$, then as $\gamma\to0$ the joint interference term in the numerators of \eqref{eq:caverages} vanishes, $\sin(\gamma/2)\cot(\gamma/2 + \phi^d) \to 0$, yielding the \emph{detector-independent} quantities,
\begin{subequations}\label{eq:wvalue}
\begin{align}
  \cmean{S_1}{\op{\sigma}^s_3} &\to \frac{\delta^s_1\,  + \delta^s_2\, }{\beta^s_+ - V^s \cos\phi^s}, \\
  \cmean{S_2}{\op{\sigma}^s_3} &\to \frac{\delta^s_1\,  - \delta^s_2\, }{\beta^s_- + V^s \cos\phi^s}.
\end{align}
\end{subequations}
In contrast with the unambiguous case, system interference remains in the denominators.

These expressions match the real parts of the \emph{weak value} expressions defined in Aharonov et al.\cite{Aharonov1988}
\begin{subequations}\label{eq:wvalueaav}
\begin{align}
  \cmean{S_1}{\op{\sigma}^s_3}_w &= \frac{\bra{S_1}\op{\sigma}^s_3\ket{\psi^s}}{\iprod{S_1}{\psi^s}}, \\
  &= \frac{\delta^s_1\,  + \delta^s_2\, }{\beta^s_+ - V^s \cos\phi^s} - i \frac{V^s \sin\phi^s}{\beta^s_+ - V^s \cos\phi^s}, \nonumber \\
  \cmean{S_2}{\op{\sigma}^s_3}_w &= \frac{\bra{S_2}\op{\sigma}^s_3\ket{\psi^s}}{\iprod{S_2}{\psi^s}}, \\
  &= \frac{\delta^s_1\,  - \delta^s_2\, }{\beta^s_+ + V^s \cos\phi^s} + i \frac{V^s \sin\phi^s}{\beta^s_+ + V^s \cos\phi^s}. \nonumber
\end{align}
\end{subequations}
As pointed out in Dressel et al.\cite{Dressel2010} any additional unitary coupling evolution such as \eqref{eq:udisturb} could in principle affect the convergence of the conditioned averages to these detector-independent weak value expressions.  However, in this case the limit is unaffected and the standard expressions are recovered. 

For the semi-weak measurement case when $\phi^d = n \pi$ with $n$ an integer, then as $\gamma\to0$ the joint interference term in the numerators of \eqref{eq:caverages} reduces to $\sin(\gamma/2)\cot(\gamma/2 + n\pi)\to(-1)^n$; therefore, system interference remains in \emph{both} the numerators and denominators, yielding the modified expressions,
\begin{subequations}\label{eq:semiwvalue}
\begin{align}
  \cmean{S_1}{\op{\sigma}^s_3} &\to \frac{\delta^s_1\,  + \delta^s_2\,  - (-1)^n V^s\cos\phi^s}{\beta^s_+ - V^s \cos\phi^s}, \\
  \cmean{S_2}{\op{\sigma}^s_3} &\to \frac{\delta^s_1\,  - \delta^s_2\,  + (-1)^n V^s\cos\phi^s}{\beta^s_- + V^s \cos\phi^s}.
\end{align}
\end{subequations}
The integer $n$ selects which detector drain is projective.  Hence, unlike the weak values \eqref{eq:wvalueaav}, these ``semi-weak values'' do not conform to general detector-independent expressions, but explicitly depend on the details of the measurement operators \eqref{eq:effmeasops}.

We emphasize that the detector tunings $\phi^d = n \pi$ are critical points around which the $\gamma\to0$ limit is \emph{unstable}, meaning that any laboratory approach to the weak coupling limit of the conditioned averages \eqref{eq:caverages} can approximate either the weak values \eqref{eq:wvalue} or the semi-weak values \eqref{eq:semiwvalue} depending on the relative magnitudes of $\gamma$ and $\phi^d$.  Hence, the limiting values will compete with each other as $\gamma$ becomes small and $\phi^d$ approaches a critical point.  The difference between these limiting cases is plotted in Figure~\ref{fig:wvalues} using the choice $n=0$.

\section{Conclusion}\label{sec:conclusion}
This work describes in detail how one electronic Mach-Zehnder interferometer may be used as a detector for the which-path information of a second Mach-Zehnder interferometer.  We use a combination of the Landauer-B\"uttiker formalism of coherent charge transport and the contextual values formalism of generalized measurement to demonstrate that inducing an interaction phase shift in the joint state for adjacent arms of the interferometers allows only the which-path information to be inferred.  For clarity of discussion, we use a simplified two-particle model of a Coulomb-induced joint phase shift in the low-bias regime to couple the interferometers and perform the measurement; however, the technique may be extended to a microscopic many-body model of the coupling interaction as well.  The efficiency of the which-path measurement depends on the interference visibility in the detector, demonstrating wave-particle complementarity between the detector and the system being measured.  Moreover, we find that the total ambiguity of the measurement depends not only on the coupling strength and the measurement efficiency, but also on the tuning phase of the detector interferometer.  Such additional measurement ambiguity leads to quantum erasure of the potentially extractable which-path information in the detector state; the erasure can be exposed by conditioning the system statistics on specific detector outcomes.

The contextual values, or generalized eigenvalues, of the which-path operator become amplified from the eigenvalues to compensate for any ambiguity in the measurement.  The analytic form of the amplification gives direct insight into the sources of the ambiguity and indicates the proper background removal for the detector.  The ambiguity compensation allows correct which-path information to be obtained on average, despite inefficiency of the measurement.  The contextual values also allow the principled construction of conditioned which-path averages for any coupling strength.  In general, such conditioned averages contain information about both the detector and the system being measured due to the joint quantum interference, but in the strong or weak coupling limits they can converge to detector-independent quantities.  In the weak coupling limit, we can obtain not only detector-independent weak values from the conditioned averages, but also detector-dependent \emph{semi-weak} values at critical points of the detector tuning.  The two distinct limits will compete depending on the relative magnitudes of the coupling strength and the detector tuning phase.

In addition to providing a simple but nontrivial theoretical example of a quantum mechanical detector interacting with an identical quantum mechanical system, the coupled electronic interferometers that we consider can be realized with current technology.  Furthermore, our results will hold with only minor modifications for an optical equivalent using a two-photon interaction (such as a nonlinear crystal) to induce the relative phase shift between copropagating interferometer arms.  Since the information being measured can be easily controlled by the experimenter in both situations, our measurement procedure can be verified experimentally.  The same techniques used here can be used to calibrate and analyze quantum mechanical detectors for less controllable system information in future experiments.

We acknowledge helpful discussions with I. Neder and support from the National Science Foundation under Grant No. DMR-0844899, and the US Army Research Office under grant Grant No. W911NF-09-0-01417.

\appendix
\section{Interaction Phase}

It may not be apparent that an extended Coulomb interaction between scattering excitations in adjacent edge states can result in an additional joint phase accumulation $\gamma$ without destroying the phase coherence.  Indeed, in momentum-space the Coulomb interaction explicitly involves energy exchange between the adjacent excitations, which would seem to imply that phase disruption would occur.  To assuage such concerns, we shall solve a simple model of Coulomb edge state interaction for the two-excitation amplitude in the co-propagating region.  We shall see that it is sufficient to keep the total energy of both excitations constant in order to obtain a joint phase accumulation over the interaction length; the fact that the excitations may exchange energy between them does not disrupt the joint phase coherence.

Consider the longitudinal part of a two-particle amplitude $\psi(x_1,x_2,t)$ that describes chiral copropagation at a speed $v$ along linear edge channels separated by a fixed distance $d$.  In the absence of any Coulomb interaction between the channels, $\psi$ must satisfy the effective Schr\"odinger equation,
\begin{align}\label{eq:appeq0}
  i \hbar \partial_t \psi = -\frac{i \hbar v}{2}(\partial_{x_1} + \partial_{x_2}) \psi.
\end{align}
For a fixed total energy $E$, the general solution of this equation has the form,
\begin{subequations}\label{eq:appsol0}
\begin{align}
  \psi_E(x_1,x_2,t) &= \psi_0(t) \psi_1(x_1,x_2) \chi(x_2 - x_1), \\
  \psi_0(t) &= \psi_0 \exp\left(-i\frac{E}{\hbar}t\right), \\
  \psi_1(x_1,x_2) &= \exp\left(\frac{iE}{\hbar v}(x_1 + x_2)\right),
\end{align}  
\end{subequations}
where $\chi$ is an arbitrary function of the difference of the coordinates.  

By choosing an initial boundary condition to be a product state of excitation scattering states at distinct energies,
\begin{align}
  E_p^\pm &= E(1\pm p),
\end{align}
where $p \in [-1,1]$ such that the energy matching condition $E_p^+ + E_p^- = 2E$ is satisfied and $E$ is correctly quantized, we fix $\chi$ to find the general product form for a fixed joint energy of $2E$,
\begin{subequations}\label{eq:appsol1}
\begin{align}
  \psi_p(2E,x_1,x_2,t) &= \xi(E_p^+,x_1,t)\xi(E_p^-,x_2,t), \\
  \xi(E,x,t) &= \xi_0 \exp\left(-i\frac{E}{\hbar}(t - \frac{2x}{v}) \right).
\end{align}
\end{subequations}
As expected, the channel amplitudes are completely decoupled in the absence of interaction and independently phase-coherent.

For a low-biased source, each single-particle energy will be approximately the Fermi energy $E_p^+ \approx E_p^- \approx E_F$, and the propagation speed will be the Fermi velocity $v = v_F$.  Therefore, in the absence of interaction each particle will accumulate a dynamical phase,
\begin{align}\label{eq:appsoldyn}
  \phi(E_F,L) = \frac{2 E_F}{\hbar v_F} L,
\end{align}
after a propagation length $L$, leading to a total joint dynamical phase of $4 E_F L / \hbar v_F$. 

If the co-propagating excitations are instead allowed to interact via a screened Coulomb potential, the effective Schr\"odinger equation must be modified to,
\begin{subequations}
\begin{align}
  i \hbar \partial_t \tilde{\psi} &= -\frac{i \hbar v}{2}(\partial_{x_1} + \partial_{x_2}) \tilde{\psi} + \frac{\alpha e^2}{r} e^{-r/\lambda}\tilde{\psi}, \\
  r &= \sqrt{d^2 + |x_2 - x_1|^2},
\end{align}
\end{subequations}
where $\lambda$ is the screening length, $r$ is the interaction distance that depends in the interacting region on the difference $|x_2-x_1|$ between the coordinates as well as the distance $d$ between the edge channels, and $\alpha$ is the Coulomb interaction constant in appropriate units.  This linear equation decouples in the coordinates $y_1 = x_1 + x_2$ and $y_2 = x_2 - x_1$, so it may still be exactly solved.  

For a fixed joint energy $E$, the general solution has the form,
\begin{subequations}\label{eq:appsol2}
\begin{align}
  \tilde{\psi}_E(x_1,x_2,t) &= \psi_0(t) \tilde{\psi}_1(x_1,x_2) \tilde{\chi}(x_2 - x_1), \\
  \tilde{\psi}_1(x_1,x_2) &= \exp\left(i \tilde{k}(E,x_1,x_2) (x_1 + x_2)\right), \\
  \label{eq:appsol2k}
  \tilde{k}(E,x_1,x_2) &= \frac{1}{\hbar v}(E - \frac{\alpha e^2}{r} e^{-r/\lambda})
\end{align}
\end{subequations}
where $\psi_0$ is the same as in \eqref{eq:appsol0} and $\tilde{\chi}$ is another arbitrary function of the difference of coordinates.  The Coulomb potential thus gives an effective position-dependent shift to the joint wave-number $\tilde{k}$ for the amplitude, which will affect the dynamical phase accumulation for the joint amplitude.

If we demand that for $r \gg \lambda$ the general solution \eqref{eq:appsol2} should reduce to the noninteracting solution \eqref{eq:appsol1}, then we find the simplest form,
\begin{subequations}\label{eq:appsol3}
\begin{align}
  \tilde{\psi}_p(x_1,x_2,t) &= \psi_p(2E,x_1,x_2,t)\, e^{- i \gamma(x_1,x_2)}, \\
  \gamma(x_1,x_2) &= \frac{\alpha e^2}{\hbar r}e^{-r/\lambda}\frac{x_1 + x_2}{v},
\end{align}
\end{subequations}
where $\psi_p$ is the noninteracting solution \eqref{eq:appsol1}.  The net effect of the Coulomb interaction between the channels is thus to contribute a position-dependent (but energy- and time-independent) phase $\gamma(x_1,x_2)$ that entangles the coordinates of the channels.  Any remaining correction factor $\tilde{\chi}(x_2 - x_1)$ to this simple form must satisfy $\tilde{\chi}(0) = 1$, so we will safely neglect it in what follows.

The excitations are collected at ohmic drains at \emph{fixed} positions $x_1 = L_1$ and $x_2 = L_2$ of the coordinates, so the detected phase $\gamma(L_1,L_2)$ will be fixed by the geometry and stable for any pair detection.  Scattering states with fixed energy such as \eqref{eq:appsol3} are stationary and extended throughout the interaction region, which explains the geometric nature of the interaction-induced phase.  The square of the wave-function $|\tilde{\psi}_p(2E,L_1,L_2,t)|^2$ indicates the (typically small) probability that the excitations will be detected \emph{simultaneously} at any particular $t$ at the drains.  However, by using a coincidence post-selection or by engineering a correlated initial scattering state, one can in principle restrict the bulk of the measured detections to be coincident.

If we further assume that before and after a co-propagation length $L$ where the channels are a fixed distance $d$ apart both edge channels rapidly split away from one another, then to good approximation the Coulomb interaction only affects the region of length $L$.  After the interaction region, the equation of motion for the state will effectively revert to \eqref{eq:appeq0}, restoring noninteracting dynamical phase accumulation similar to \eqref{eq:appsoldyn}.  Hence, the amplitude for jointly detecting the excitations at ohmic drain positions $L_1 > L$ and $L_2 > L$ will contain an extra dynamical phase that is accumulated by the joint state only within the interaction region,
\begin{subequations}\label{eq:appsol4}
\begin{align}
  \tilde{\psi}_p(L_1,L_2,t) &= \psi_p(2E,L_1,L_2,t)\, e^{- i \gamma}, \\
  \gamma = \gamma(L,L) &= \frac{\alpha e^2}{\hbar d}e^{-d/\lambda}\frac{2L}{v}.
\end{align}
\end{subequations}
Moreover, the joint state may be further scattered after accumulating the joint interaction phase but before the joint detection, as indicated in the main text.  We see that for such simultaneous detection the interaction phase $\gamma$ will be linear in the interaction length $L$ and therefore should be tunable in principle.  

If the detections are not simultaneous, then one excitation will be detected in a drain at a time $t_1$, followed by the second excitation at a later time $t_2$.  The joint state will therefore be successively collapsed by the drain detections, which will introduce an additional relative phase factor due to the discrepancy in detection time.  Specifically, for $x_1,x_2 > L$ the joint state will have the form,
\begin{align}
  \tilde{\psi}_p(x_1,x_2,t) &= \psi_p(2E,x_1,x_2,t)\, e^{- i \gamma},
\end{align}
with the accumulated interaction phase $\gamma$ as in \eqref{eq:appsol4}.  Detection of the first excitation at $L_1$ at $t_1$ then collapses the joint state to,
\begin{subequations}
\begin{align}
  \tilde{\psi}'_p(x_2,t) &= \frac{\psi_p(2E,L_1,x_2,t)}{\sqrt{P(L_1,t_1)}}\, e^{- i \gamma}, \\
  P(L_1,t_1) &= \int dx_2\, |\tilde{\psi}_p(L_1,x_2,t_1)|^2.
\end{align}
\end{subequations}
Evolving the remaining single particle state to time $t_2$ and then detecting the second excitation at $L_2$ produces the amplitude, 
\begin{align}
  \tilde{\psi}''_p &= \frac{\psi_p(2E,L_1,L_2,t_1)}{\sqrt{P(L_1,t_1)}}\, e^{- i (\gamma + E (t_2 - t_1)/\hbar)}.
\end{align}
The only difference between the sequential detection amplitudes and the joint detection amplitude \eqref{eq:appsol4} is the extra temporal phase that is accumulated between the two detections.  Notably, the extra temporal phase factor appears as a global phase that should not affect the final statistics, in contrast to the geometric interaction phase $\gamma$, which can be exposed by further scattering before the sequential detection as in the main text.  Allowing for fluctuations in $\gamma$ as in \S\ref{sec:fluctuation} will account for geometric uncertainty in the interaction length, as well as the pair injection frequency of the source.


\begin{thebibliography}{10}%
\makeatletter
\providecommand \@ifxundefined [1]{%
 \ifx #1\undefined \expandafter \@firstoftwo
 \else \expandafter \@secondoftwo
\fi
}%
\providecommand \@ifnum [1]{%
 \ifnum #1\expandafter \@firstoftwo
 \else \expandafter \@secondoftwo
\fi
}%
\providecommand \enquote [1]{``#1''}%
\providecommand \bibnamefont  [1]{#1}%
\providecommand \bibfnamefont [1]{#1}%
\providecommand \citenamefont [1]{#1}%
\providecommand\href[0]{\@sanitize\@href}%
\providecommand\@href[1]{\endgroup\@@startlink{#1}\endgroup\@@href}%
\providecommand\@@href[1]{#1\@@endlink}%
\providecommand \@sanitize [0]{\begingroup\catcode`\&12\catcode`\#12\relax}%
\@ifxundefined \pdfoutput {\@firstoftwo}{%
 \@ifnum{\z@=\pdfoutput}{\@firstoftwo}{\@secondoftwo}%
}{%
 \providecommand\@@startlink[1]{\leavevmode}%
 \providecommand\@@endlink[0]{}%
}{%
 \providecommand\@@startlink[1]{%
  \leavevmode
  \pdfstartlink
   attr{/Border[0 0 1 ]/H/I/C[0 1 1]}%
   user{/Subtype/Link/A<</Type/Action/S/URI/URI(#1)>>}%
  \relax
 }%
 \providecommand\@@endlink[0]{\pdfendlink}%
}%
\providecommand \url  [0]{\begingroup\@sanitize \@url }%
\providecommand \@url [1]{\endgroup\@href {#1}{\urlprefix}}%
\providecommand \urlprefix [0]{URL }%
\providecommand \Eprint[0]{\href }%
\@ifxundefined \urlstyle {%
  \providecommand \doi [1]{doi:\discretionary{}{}{}#1}%
}{%
  \providecommand \doi [0]{doi:\discretionary{}{}{}\begingroup
  \urlstyle{rm}\Url }%
}%
\providecommand \doibase [0]{http://dx.doi.org/}%
\providecommand \Doi[1]{\href{\doibase#1}}%
\providecommand \bibAnnote [3]{%
  \BibitemShut{#1}%
  \begin{quotation}\noindent
    \textsc{Key:}\ #2\\\textsc{Annotation:}\ #3%
  \end{quotation}%
}%
\providecommand \bibAnnoteFile [2]{%
  \IfFileExists{#2}{\bibAnnote {#1} {#2} {\input{#2}}}{}%
}%
\providecommand \typeout [0]{\immediate \write \m@ne }%
\providecommand \selectlanguage [0]{\@gobble}%
\providecommand \bibinfo [0]{\@secondoftwo}%
\providecommand \bibfield [0]{\@secondoftwo}%
\providecommand \translation [1]{[#1]}%
\providecommand \BibitemOpen[0]{}%
\providecommand \bibitemStop [0]{}%
\providecommand \bibitemNoStop [0]{.\EOS\space}%
\providecommand \EOS [0]{\spacefactor3000\relax}%
\providecommand \BibitemShut [1]{\csname bibitem#1\endcsname}%
\bibitem{Ji2003}%
  \BibitemOpen
  \bibfield{author}{%
  \bibinfo {author} {\bibfnamefont{Y.}~\bibnamefont{Ji}}, \bibinfo {author}
  {\bibfnamefont{Y.}~\bibnamefont{Chung}}, \bibinfo {author}
  {\bibfnamefont{D.}~\bibnamefont{Sprinzak}}, \bibinfo {author}
  {\bibfnamefont{M.}~\bibnamefont{Heiblum}}, \bibinfo {author}
  {\bibfnamefont{D.}~\bibnamefont{Mahalu}},\ and\ \bibinfo {author}
  {\bibfnamefont{H.}~\bibnamefont{Shtrikman}},\ }%
  \bibfield{journal}{%
  \bibinfo {journal} {Nature}\ }%
  \textbf{\bibinfo {volume} {422}},\ \bibinfo {pages} {415} (\bibinfo {year}
  {2003})%
  \bibAnnoteFile{NoStop}{Ji2003}%
\bibitem{Devyatov2007}%
  \BibitemOpen
  \bibfield{author}{%
  \bibinfo {author} {\bibfnamefont{E.~V.}\ \bibnamefont{Devyatov}},\ }%
  \bibfield{journal}{%
  \bibinfo {journal} {Physics-Uspekhi}\ }%
  \textbf{\bibinfo {volume} {50}},\ \bibinfo {pages} {197} (\bibinfo {year}
  {2007})%
  \bibAnnoteFile{NoStop}{Devyatov2007}%
\bibitem{Buttiker1992}%
  \BibitemOpen
  \bibfield{author}{%
  \bibinfo {author} {\bibfnamefont{M.}~\bibnamefont{B\"uttiker}},\ }%
  \bibfield{journal}{%
  \bibinfo {journal} {Phys. Rev. B}\ }%
  \textbf{\bibinfo {volume} {46}},\ \bibinfo {pages} {12485 } (\bibinfo {year}
  {1992})%
  \bibAnnoteFile{NoStop}{Buttiker1992}%
\bibitem{Martin1992}%
  \BibitemOpen
  \bibfield{author}{%
  \bibinfo {author} {\bibfnamefont{T.}~\bibnamefont{Martin}}\ and\ \bibinfo
  {author} {\bibfnamefont{R.}~\bibnamefont{Landauer}},\ }%
  \bibfield{journal}{%
  \bibinfo {journal} {Phys. Rev. B}\ }%
  \textbf{\bibinfo {volume} {45}},\ \bibinfo {pages} {1742 } (\bibinfo {year}
  {1992})%
  \bibAnnoteFile{NoStop}{Martin1992}%
\bibitem{Blanter2000}%
  \BibitemOpen
  \bibfield{author}{%
  \bibinfo {author} {\bibfnamefont{Y.}~\bibnamefont{Blanter}}\ and\ \bibinfo
  {author} {\bibfnamefont{M.}~\bibnamefont{B\"{u}ttiker}},\ }%
  \bibfield{journal}{%
  \bibinfo {journal} {Phys. Rep.}\ }%
  \textbf{\bibinfo {volume} {336}} (\bibinfo {year} {2000})%
  \bibAnnoteFile{NoStop}{Blanter2000}%
\bibitem{Neder2007b}%
  \BibitemOpen
  \bibfield{author}{%
  \bibinfo {author} {\bibfnamefont{I.}~\bibnamefont{Neder}}, \bibinfo {author}
  {\bibfnamefont{N.}~\bibnamefont{Ofek}}, \bibinfo {author}
  {\bibfnamefont{Y.}~\bibnamefont{Chung}}, \bibinfo {author}
  {\bibfnamefont{M.}~\bibnamefont{Heiblum}}, \bibinfo {author}
  {\bibfnamefont{D.}~\bibnamefont{Mahalu}},\ and\ \bibinfo {author}
  {\bibfnamefont{V.}~\bibnamefont{Umansky}},\ }%
  \bibfield{journal}{%
  \bibinfo {journal} {Nature}\ }%
  \textbf{\bibinfo {volume} {448}},\ \bibinfo {pages} {333} (\bibinfo {year}
  {2007})%
  \bibAnnoteFile{NoStop}{Neder2007b}%
\bibitem{Giovannetti2008}%
  \BibitemOpen
  \bibfield{author}{%
  \bibinfo {author} {\bibfnamefont{V.}~\bibnamefont{Giovannetti}}, \bibinfo
  {author} {\bibfnamefont{F.}~\bibnamefont{Taddei}}, \bibinfo {author}
  {\bibfnamefont{D.}~\bibnamefont{Frustaglia}},\ and\ \bibinfo {author}
  {\bibfnamefont{R.}~\bibnamefont{Fazio}},\ }%
  \bibfield{journal}{%
  \bibinfo {journal} {Phys. Rev. B}\ }%
  \textbf{\bibinfo {volume} {77}},\ \bibinfo {pages} {155320} (\bibinfo {year}
  {2008})%
  \bibAnnoteFile{NoStop}{Giovannetti2008}%
\bibitem{Lin2009}%
  \BibitemOpen
  \bibfield{author}{%
  \bibinfo {author} {\bibfnamefont{P.~V.}\ \bibnamefont{Lin}}, \bibinfo
  {author} {\bibfnamefont{F.~E.}\ \bibnamefont{Camino}},\ and\ \bibinfo
  {author} {\bibfnamefont{V.~J.}\ \bibnamefont{Goldman}},\ }%
  \bibfield{journal}{%
  \bibinfo {journal} {Phys. Rev. B}\ }%
  \textbf{\bibinfo {volume} {80}},\ \bibinfo {pages} {125310} (\bibinfo {year}
  {2009})%
  \bibAnnoteFile{NoStop}{Lin2009}%
\bibitem{Chirolli2010}%
  \BibitemOpen
  \bibfield{author}{%
  \bibinfo {author} {\bibfnamefont{L.}~\bibnamefont{Chirolli}}, \bibinfo
  {author} {\bibfnamefont{E.}~\bibnamefont{Strambini}}, \bibinfo {author}
  {\bibfnamefont{V.}~\bibnamefont{Giovannetti}}, \bibinfo {author}
  {\bibfnamefont{F.}~\bibnamefont{Taddei}}, \bibinfo {author}
  {\bibfnamefont{V.}~\bibnamefont{Piazza}}, \bibinfo {author}
  {\bibfnamefont{R.}~\bibnamefont{Fazio}}, \bibinfo {author}
  {\bibfnamefont{F.}~\bibnamefont{Beltram}},\ and\ \bibinfo {author}
  {\bibfnamefont{G.}~\bibnamefont{Burkard}},\ }%
  \bibfield{journal}{%
  \bibinfo {journal} {Phys. Rev. B}\ }%
  \textbf{\bibinfo {volume} {82}},\ \bibinfo {pages} {045403} (\bibinfo {year}
  {2010})%
  \bibAnnoteFile{NoStop}{Chirolli2010}%
\bibitem{Aharonov1959}%
  \BibitemOpen
  \bibfield{author}{%
  \bibinfo {author} {\bibfnamefont{Y.}~\bibnamefont{Aharonov}}\ and\ \bibinfo
  {author} {\bibfnamefont{D.}~\bibnamefont{Bohm}},\ }%
  \bibfield{journal}{%
  \bibinfo {journal} {Phys. Rev.}\ }%
  \textbf{\bibinfo {volume} {115}},\ \bibinfo {pages} {485 } (\bibinfo {year}
  {1959})%
  \bibAnnoteFile{NoStop}{Aharonov1959}%
\bibitem{Anandan1992}%
  \BibitemOpen
  \bibfield{author}{%
  \bibinfo {author} {\bibfnamefont{J.}~\bibnamefont{Anandan}},\ }%
  \bibfield{journal}{%
  \bibinfo {journal} {Nature}\ }%
  \textbf{\bibinfo {volume} {360}},\ \bibinfo {pages} {307} (\bibinfo {year}
  {1992})%
  \bibAnnoteFile{NoStop}{Anandan1992}%
\bibitem{Chung2005}%
  \BibitemOpen
  \bibfield{author}{%
  \bibinfo {author} {\bibfnamefont{V.~S.-W.}\ \bibnamefont{Chung}}, \bibinfo
  {author} {\bibfnamefont{P.}~\bibnamefont{Samuelsson}},\ and\ \bibinfo
  {author} {\bibfnamefont{M.}~\bibnamefont{B\"uttiker}},\ }%
  \bibfield{journal}{%
  \bibinfo {journal} {Phys. Rev. B}\ }%
  \textbf{\bibinfo {volume} {72}},\ \bibinfo {pages} {125320} (\bibinfo {year}
  {2005})%
  \bibAnnoteFile{NoStop}{Chung2005}%
\bibitem{Forster2005}%
  \BibitemOpen
  \bibfield{author}{%
  \bibinfo {author} {\bibfnamefont{H.}~\bibnamefont{F\"{o}rster}}, \bibinfo
  {author} {\bibfnamefont{S.}~\bibnamefont{Pilgram}},\ and\ \bibinfo {author}
  {\bibfnamefont{M.}~\bibnamefont{B\"{u}ttiker}},\ }%
  \bibfield{journal}{%
  \bibinfo {journal} {Phys. Rev. B}\ }%
  \textbf{\bibinfo {volume} {72}},\ \bibinfo {pages} {075301} (\bibinfo {year}
  {2005})%
  \bibAnnoteFile{NoStop}{Forster2005}%
\bibitem{Chalker2007}%
  \BibitemOpen
  \bibfield{author}{%
  \bibinfo {author} {\bibfnamefont{J.~T.}\ \bibnamefont{Chalker}}, \bibinfo
  {author} {\bibfnamefont{Y.}~\bibnamefont{Gefen}},\ and\ \bibinfo {author}
  {\bibfnamefont{M.~Y.}\ \bibnamefont{Veillette}},\ }%
  \bibfield{journal}{%
  \bibinfo {journal} {Phys. Rev. B}\ }%
  \textbf{\bibinfo {volume} {76}},\ \bibinfo {pages} {085320} (\bibinfo {year}
  {2007})%
  \bibAnnoteFile{NoStop}{Chalker2007}%
\bibitem{Litvin2007}%
  \BibitemOpen
  \bibfield{author}{%
  \bibinfo {author} {\bibfnamefont{L.~V.}\ \bibnamefont{Litvin}}, \bibinfo
  {author} {\bibfnamefont{H.-P.}\ \bibnamefont{Tranitz}}, \bibinfo {author}
  {\bibfnamefont{W.}~\bibnamefont{Wegscheider}},\ and\ \bibinfo {author}
  {\bibfnamefont{C.}~\bibnamefont{Strunk}},\ }%
  \bibfield{journal}{%
  \bibinfo {journal} {Phys. Rev. B}\ }%
  \textbf{\bibinfo {volume} {75}},\ \bibinfo {pages} {33315} (\bibinfo {year}
  {2007})%
  \bibAnnoteFile{NoStop}{Litvin2007}%
\bibitem{Neder2007a}%
  \BibitemOpen
  \bibfield{author}{%
  \bibinfo {author} {\bibfnamefont{I.}~\bibnamefont{Neder}}, \bibinfo {author}
  {\bibfnamefont{M.}~\bibnamefont{Heiblum}}, \bibinfo {author}
  {\bibfnamefont{D.}~\bibnamefont{Mahalu}},\ and\ \bibinfo {author}
  {\bibfnamefont{V.}~\bibnamefont{Umansky}},\ }%
  \bibfield{journal}{%
  \bibinfo {journal} {Phys. Rev. Lett.}\ }%
  \textbf{\bibinfo {volume} {98}},\ \bibinfo {pages} {36803} (\bibinfo {year}
  {2007})%
  \bibAnnoteFile{NoStop}{Neder2007a}%
\bibitem{Roulleau2008}%
  \BibitemOpen
  \bibfield{author}{%
  \bibinfo {author} {\bibfnamefont{P.}~\bibnamefont{Roulleau}}, \bibinfo
  {author} {\bibfnamefont{F.}~\bibnamefont{Portier}}, \bibinfo {author}
  {\bibfnamefont{D.~C.}\ \bibnamefont{Glattli}}, \bibinfo {author}
  {\bibfnamefont{P.}~\bibnamefont{Roche}}, \bibinfo {author}
  {\bibfnamefont{A.}~\bibnamefont{Cavanna}}, \bibinfo {author}
  {\bibfnamefont{G.}~\bibnamefont{Faini}}, \bibinfo {author}
  {\bibfnamefont{U.}~\bibnamefont{Gennser}},\ and\ \bibinfo {author}
  {\bibfnamefont{D.}~\bibnamefont{Mailly}},\ }%
  \bibfield{journal}{%
  \bibinfo {journal} {Phys. Rev. Lett.}\ }%
  \textbf{\bibinfo {volume} {100}},\ \bibinfo {pages} {126802} (\bibinfo {year}
  {2008})%
  \bibAnnoteFile{NoStop}{Roulleau2008}%
\bibitem{Youn2009}%
  \BibitemOpen
  \bibfield{author}{%
  \bibinfo {author} {\bibfnamefont{S.-C.}\ \bibnamefont{Youn}}, \bibinfo
  {author} {\bibfnamefont{H.-W.}\ \bibnamefont{Lee}},\ and\ \bibinfo {author}
  {\bibfnamefont{H.-S.}\ \bibnamefont{Sim}},\ }%
  \bibfield{journal}{%
  \bibinfo {journal} {Phys. Rev. B}\ }%
  \textbf{\bibinfo {volume} {80}},\ \bibinfo {pages} {113307} (\bibinfo {year}
  {2009})%
  \bibAnnoteFile{NoStop}{Youn2009}%
\bibitem{Hashisaka2010}%
  \BibitemOpen
  \bibfield{author}{%
  \bibinfo {author} {\bibfnamefont{M.}~\bibnamefont{Hashisaka}}, \bibinfo
  {author} {\bibfnamefont{A.}~\bibnamefont{Helzel}}, \bibinfo {author}
  {\bibfnamefont{S.}~\bibnamefont{Nakamura}}, \bibinfo {author}
  {\bibfnamefont{L.}~\bibnamefont{Litvin}}, \bibinfo {author}
  {\bibfnamefont{Y.}~\bibnamefont{Yamauchi}}, \bibinfo {author}
  {\bibfnamefont{K.}~\bibnamefont{Kobayashi}}, \bibinfo {author}
  {\bibfnamefont{T.}~\bibnamefont{Ono}}, \bibinfo {author}
  {\bibfnamefont{H.-P.}\ \bibnamefont{Tranitz}}, \bibinfo {author}
  {\bibfnamefont{W.}~\bibnamefont{Wegscheider}},\ and\ \bibinfo {author}
  {\bibfnamefont{C.}~\bibnamefont{Strunk}},\ }%
  \bibfield{journal}{%
  \bibinfo {journal} {Physica E}\ }%
  \textbf{\bibinfo {volume} {42}},\ \bibinfo {pages} {1091} (\bibinfo {year}
  {2010})%
  \bibAnnoteFile{NoStop}{Hashisaka2010}%
\bibitem{Khym2009}%
  \BibitemOpen
  \bibfield{author}{%
  \bibinfo {author} {\bibfnamefont{G.~L.}\ \bibnamefont{Khym}}\ and\ \bibinfo
  {author} {\bibfnamefont{K.}~\bibnamefont{Kang}},\ }%
  \bibfield{journal}{%
  \bibinfo {journal} {Phys. Rev. B}\ }%
  \textbf{\bibinfo {volume} {79}},\ \bibinfo {pages} {195306} (\bibinfo {year}
  {2009})%
  \bibAnnoteFile{NoStop}{Khym2009}%
\bibitem{Haack2010}%
  \BibitemOpen
  \bibfield{author}{%
  \bibinfo {author} {\bibfnamefont{G.}~\bibnamefont{Haack}}, \bibinfo {author}
  {\bibfnamefont{H.}~\bibnamefont{F\"{o}rster}},\ and\ \bibinfo {author}
  {\bibfnamefont{M.}~\bibnamefont{B\"{u}ttiker}},\ }%
  \bibfield{journal}{%
  \bibinfo {journal} {Phys. Rev. B}\ }%
  \textbf{\bibinfo {volume} {82}},\ \bibinfo {pages} {155303} (\bibinfo {year}
  {2010})%
  \bibAnnoteFile{NoStop}{Haack2010}%
\bibitem{Neder2006}%
  \BibitemOpen
  \bibfield{author}{%
  \bibinfo {author} {\bibfnamefont{I.}~\bibnamefont{Neder}}, \bibinfo {author}
  {\bibfnamefont{M.}~\bibnamefont{Heiblum}}, \bibinfo {author}
  {\bibfnamefont{Y.}~\bibnamefont{Levinson}}, \bibinfo {author}
  {\bibfnamefont{D.}~\bibnamefont{Mahalu}},\ and\ \bibinfo {author}
  {\bibfnamefont{V.}~\bibnamefont{Umansky}},\ }%
  \bibfield{journal}{%
  \bibinfo {journal} {Phys. Rev. Lett.}\ }%
  \textbf{\bibinfo {volume} {96}},\ \bibinfo {pages} {16804} (\bibinfo {year}
  {2006})%
  \bibAnnoteFile{NoStop}{Neder2006}%
\bibitem{Roulleau2007}%
  \BibitemOpen
  \bibfield{author}{%
  \bibinfo {author} {\bibfnamefont{P.}~\bibnamefont{Roulleau}}, \bibinfo
  {author} {\bibfnamefont{F.}~\bibnamefont{Portier}}, \bibinfo {author}
  {\bibfnamefont{D.~C.}\ \bibnamefont{Glattli}}, \bibinfo {author}
  {\bibfnamefont{P.}~\bibnamefont{Roche}}, \bibinfo {author}
  {\bibfnamefont{A.}~\bibnamefont{Cavanna}}, \bibinfo {author}
  {\bibfnamefont{G.}~\bibnamefont{Faini}}, \bibinfo {author}
  {\bibfnamefont{U.}~\bibnamefont{Gennser}},\ and\ \bibinfo {author}
  {\bibfnamefont{D.}~\bibnamefont{Mailly}},\ }%
  \bibfield{journal}{%
  \bibinfo {journal} {Phy. Rev. B}\ }%
  \textbf{\bibinfo {volume} {76}},\ \bibinfo {pages} {161309} (\bibinfo {year}
  {2007})%
  \bibAnnoteFile{NoStop}{Roulleau2007}%
\bibitem{Litvin2008}%
  \BibitemOpen
  \bibfield{author}{%
  \bibinfo {author} {\bibfnamefont{L.~V.}\ \bibnamefont{Litvin}}, \bibinfo
  {author} {\bibfnamefont{A.}~\bibnamefont{Helzel}}, \bibinfo {author}
  {\bibfnamefont{H.~P.}\ \bibnamefont{Tranitz}}, \bibinfo {author}
  {\bibfnamefont{W.}~\bibnamefont{Wegscheider}},\ and\ \bibinfo {author}
  {\bibfnamefont{C.}~\bibnamefont{Strunk}},\ }%
  \bibfield{journal}{%
  \bibinfo {journal} {Phys. Rev. B}\ }%
  \textbf{\bibinfo {volume} {78}},\ \bibinfo {pages} {075303} (\bibinfo {year}
  {2008})%
  \bibAnnoteFile{NoStop}{Litvin2008}%
\bibitem{Bieri2009}%
  \BibitemOpen
  \bibfield{author}{%
  \bibinfo {author} {\bibfnamefont{E.}~\bibnamefont{Bieri}}, \bibinfo {author}
  {\bibfnamefont{M.}~\bibnamefont{Weiss}}, \bibinfo {author}
  {\bibfnamefont{O.}~\bibnamefont{G\"{o}ktas}}, \bibinfo {author}
  {\bibfnamefont{M.}~\bibnamefont{Hauser}}, \bibinfo {author}
  {\bibfnamefont{C.}~\bibnamefont{Sch\"{o}nenberger}},\ and\ \bibinfo {author}
  {\bibfnamefont{S.}~\bibnamefont{Oberholzer}},\ }%
  \bibfield{journal}{%
  \bibinfo {journal} {Phys. Rev. B}\ }%
  \textbf{\bibinfo {volume} {79}},\ \bibinfo {pages} {245324} (\bibinfo {year}
  {2009})%
  \bibAnnoteFile{NoStop}{Bieri2009}%
\bibitem{Sukhorukov2007a}%
  \BibitemOpen
  \bibfield{author}{%
  \bibinfo {author} {\bibfnamefont{E.~V.}\ \bibnamefont{Sukhorukov}}\ and\
  \bibinfo {author} {\bibfnamefont{V.~V.}\ \bibnamefont{Cheianov}},\ }%
  \bibfield{journal}{%
  \bibinfo {journal} {Phys. Rev. Lett.}\ }%
  \textbf{\bibinfo {volume} {99}},\ \bibinfo {pages} {156801} (\bibinfo {year}
  {2007})%
  \bibAnnoteFile{NoStop}{Sukhorukov2007a}%
\bibitem{Levkivskyi2008}%
  \BibitemOpen
  \bibfield{author}{%
  \bibinfo {author} {\bibfnamefont{I.~P.}\ \bibnamefont{Levkivskyi}}\ and\
  \bibinfo {author} {\bibfnamefont{E.~V.}\ \bibnamefont{Sukhorukov}},\ }%
  \bibfield{journal}{%
  \bibinfo {journal} {Phys. Rev. B}\ }%
  \textbf{\bibinfo {volume} {78}},\ \bibinfo {pages} {045322} (\bibinfo {year}
  {2008})%
  \bibAnnoteFile{NoStop}{Levkivskyi2008}%
\bibitem{Neder2008}%
  \BibitemOpen
  \bibfield{author}{%
  \bibinfo {author} {\bibfnamefont{I.}~\bibnamefont{Neder}}\ and\ \bibinfo
  {author} {\bibfnamefont{E.}~\bibnamefont{Ginossar}},\ }%
  \bibfield{journal}{%
  \bibinfo {journal} {Phys. Rev. Lett.}\ }%
  \textbf{\bibinfo {volume} {100}},\ \bibinfo {pages} {196806} (\bibinfo {year}
  {2008})%
  \bibAnnoteFile{NoStop}{Neder2008}%
\bibitem{Neuenhahn2008}%
  \BibitemOpen
  \bibfield{author}{%
  \bibinfo {author} {\bibfnamefont{C.}~\bibnamefont{Neuenhahn}}\ and\ \bibinfo
  {author} {\bibfnamefont{F.}~\bibnamefont{Marquardt}},\ }%
  \bibfield{journal}{%
  \bibinfo {journal} {New J. Phys.}\ }%
  \textbf{\bibinfo {volume} {10}},\ \bibinfo {pages} {115018} (\bibinfo {year}
  {2008})%
  \bibAnnoteFile{NoStop}{Neuenhahn2008}%
\bibitem{Youn2008}%
  \BibitemOpen
  \bibfield{author}{%
  \bibinfo {author} {\bibfnamefont{S.-C.}\ \bibnamefont{Youn}}, \bibinfo
  {author} {\bibfnamefont{H.-W.}\ \bibnamefont{Lee}},\ and\ \bibinfo {author}
  {\bibfnamefont{H.-S.}\ \bibnamefont{Sim}},\ }%
  \bibfield{journal}{%
  \bibinfo {journal} {Phys. Rev. Lett.}\ }%
  \textbf{\bibinfo {volume} {100}},\ \bibinfo {pages} {196807} (\bibinfo {year}
  {2008})%
  \bibAnnoteFile{NoStop}{Youn2008}%
\bibitem{Levkivskyi2009b}%
  \BibitemOpen
  \bibfield{author}{%
  \bibinfo {author} {\bibfnamefont{I.~P.}\ \bibnamefont{Levkivskyi}}\ and\
  \bibinfo {author} {\bibfnamefont{E.~V.}\ \bibnamefont{Sukhorukov}},\ }%
  \bibfield{journal}{%
  \bibinfo {journal} {Phys. Rev. Lett.}\ }%
  \textbf{\bibinfo {volume} {103}},\ \bibinfo {pages} {036801} (\bibinfo {year}
  {2009})%
  \bibAnnoteFile{NoStop}{Levkivskyi2009b}%
\bibitem{Kovrizhin2009}%
  \BibitemOpen
  \bibfield{author}{%
  \bibinfo {author} {\bibfnamefont{D.~L.}\ \bibnamefont{Kovrizhin}}\ and\
  \bibinfo {author} {\bibfnamefont{J.~T.}\ \bibnamefont{Chalker}},\ }%
  \bibfield{journal}{%
  \bibinfo {journal} {Phys. Rev. B}\ }%
  \textbf{\bibinfo {volume} {80}},\ \bibinfo {pages} {161306} (\bibinfo {year}
  {2009})%
  \bibAnnoteFile{NoStop}{Kovrizhin2009}%
\bibitem{Kovrizhin2010}%
  \BibitemOpen
  \bibfield{author}{%
  \bibinfo {author} {\bibfnamefont{D.~L.}\ \bibnamefont{Kovrizhin}}\ and\
  \bibinfo {author} {\bibfnamefont{J.~T.}\ \bibnamefont{Chalker}},\ }%
  \bibfield{journal}{%
  \bibinfo {journal} {Phys. Rev. B}\ }%
  \textbf{\bibinfo {volume} {81}},\ \bibinfo {pages} {155318} (\bibinfo {year}
  {2010})%
  \bibAnnoteFile{NoStop}{Kovrizhin2010}%
\bibitem{Hardy1992}%
  \BibitemOpen
  \bibfield{author}{%
  \bibinfo {author} {\bibfnamefont{L.}~\bibnamefont{Hardy}},\ }%
  \bibfield{journal}{%
  \bibinfo {journal} {Phys. Rev. Lett.}\ }%
  \textbf{\bibinfo {volume} {68}},\ \bibinfo {pages} {2981} (\bibinfo {year}
  {1992})%
  \bibAnnoteFile{NoStop}{Hardy1992}%
\bibitem{Aharonov2001}%
  \BibitemOpen
  \bibfield{author}{%
  \bibinfo {author} {\bibfnamefont{Y.}~\bibnamefont{Aharonov}}, \bibinfo
  {author} {\bibfnamefont{A.}~\bibnamefont{Botero}}, \bibinfo {author}
  {\bibfnamefont{S.}~\bibnamefont{Popescu}}, \bibinfo {author}
  {\bibfnamefont{B.}~\bibnamefont{Reznik}},\ and\ \bibinfo {author}
  {\bibfnamefont{J.}~\bibnamefont{Tollaksen}},\ }%
  \bibfield{journal}{%
  \bibinfo {journal} {Phys. Lett. A}\ }%
  \textbf{\bibinfo {volume} {301}},\ \bibinfo {pages} {7} (\bibinfo {year}
  {2001})%
  \bibAnnoteFile{NoStop}{Aharonov2001}%
\bibitem{Lundeen2009}%
  \BibitemOpen
  \bibfield{author}{%
  \bibinfo {author} {\bibfnamefont{J.~S.}\ \bibnamefont{Lundeen}}\ and\
  \bibinfo {author} {\bibfnamefont{A.~M.}\ \bibnamefont{Steinberg}},\ }%
  \bibfield{journal}{%
  \bibinfo {journal} {Phys. Rev. Lett.}\ }%
  \textbf{\bibinfo {volume} {102}},\ \bibinfo {pages} {020404} (\bibinfo {year}
  {2009})%
  \bibAnnoteFile{NoStop}{Lundeen2009}%
\bibitem{Yokota2009}%
  \BibitemOpen
  \bibfield{author}{%
  \bibinfo {author} {\bibfnamefont{K.}~\bibnamefont{Yokota}}, \bibinfo {author}
  {\bibfnamefont{T.}~\bibnamefont{Yamamoto}}, \bibinfo {author}
  {\bibfnamefont{M.}~\bibnamefont{Koashi}},\ and\ \bibinfo {author}
  {\bibfnamefont{N.}~\bibnamefont{Imoto}},\ }%
  \bibfield{journal}{%
  \bibinfo {journal} {New J. Phys.}\ }%
  \textbf{\bibinfo {volume} {11}},\ \bibinfo {pages} {033011} (\bibinfo {year}
  {2009})%
  \bibAnnoteFile{NoStop}{Yokota2009}%
\bibitem{Kang2007}%
  \BibitemOpen
  \bibfield{author}{%
  \bibinfo {author} {\bibfnamefont{K.}~\bibnamefont{Kang}},\ }%
  \bibfield{journal}{%
  \bibinfo {journal} {Phys. Rev. B}\ }%
  \textbf{\bibinfo {volume} {75}},\ \bibinfo {pages} {125326} (\bibinfo {year}
  {2007})%
  \bibAnnoteFile{NoStop}{Kang2007}%
\bibitem{Kang2008}%
  \BibitemOpen
  \bibfield{author}{%
  \bibinfo {author} {\bibfnamefont{K.}~\bibnamefont{Kang}}\ and\ \bibinfo
  {author} {\bibfnamefont{K.}~\bibnamefont{{Ho Lee}}},\ }%
  \bibfield{journal}{%
  \bibinfo {journal} {Physica E}\ }%
  \textbf{\bibinfo {volume} {40}},\ \bibinfo {pages} {1395} (\bibinfo {year}
  {2008})%
  \bibAnnoteFile{NoStop}{Kang2008}%
\bibitem{Braun2008}%
  \BibitemOpen
  \bibfield{author}{%
  \bibinfo {author} {\bibfnamefont{D.}~\bibnamefont{Braun}}\ and\ \bibinfo
  {author} {\bibfnamefont{M.-S.}\ \bibnamefont{Choi}},\ }%
  \bibfield{journal}{%
  \bibinfo {journal} {Phys. Rev. A}\ }%
  \textbf{\bibinfo {volume} {78}},\ \bibinfo {pages} {032114} (\bibinfo {year}
  {2008})%
  \bibAnnoteFile{NoStop}{Braun2008}%
\bibitem{Shpitalnik2008}%
  \BibitemOpen
  \bibfield{author}{%
  \bibinfo {author} {\bibfnamefont{V.}~\bibnamefont{Shpitalnik}}, \bibinfo
  {author} {\bibfnamefont{Y.}~\bibnamefont{Gefen}},\ and\ \bibinfo {author}
  {\bibfnamefont{A.}~\bibnamefont{Romito}},\ }%
  \bibfield{journal}{%
  \bibinfo {journal} {Phys. Rev. Lett.}\ }%
  \textbf{\bibinfo {volume} {101}},\ \bibinfo {pages} {226802} (\bibinfo {year}
  {2008})%
  \bibAnnoteFile{NoStop}{Shpitalnik2008}%
\bibitem{Wheeler1979}%
  \BibitemOpen
  \bibfield{author}{%
  \bibinfo {author} {\bibfnamefont{J.~A.}\ \bibnamefont{Wheeler}},\ }%
  \emph{\bibinfo {title} {{Problems in the Formulation of Physics}}}\ (\bibinfo
  {publisher} {{North Holland, Amsterdam}},\ \bibinfo {year} {1979})%
  \bibAnnoteFile{NoStop}{Wheeler1979}%
\bibitem{Scully1982}%
  \BibitemOpen
  \bibfield{author}{%
  \bibinfo {author} {\bibfnamefont{M.~O.}\ \bibnamefont{Scully}}\ and\ \bibinfo
  {author} {\bibfnamefont{K.}~\bibnamefont{Dr\"uhl}},\ }%
  \bibfield{journal}{%
  \bibinfo {journal} {Phys. Rev. A}\ }%
  \textbf{\bibinfo {volume} {25}},\ \bibinfo {pages} {2208} (\bibinfo {year}
  {1982})%
  \bibAnnoteFile{NoStop}{Scully1982}%
\bibitem{Scully1991}%
  \BibitemOpen
  \bibfield{author}{%
  \bibinfo {author} {\bibfnamefont{M.}~\bibnamefont{Scully}}, \bibinfo {author}
  {\bibfnamefont{B.}~\bibnamefont{Englert}},\ and\ \bibinfo {author}
  {\bibfnamefont{H.}~\bibnamefont{Walther}},\ }%
  \bibfield{journal}{%
  \bibinfo {journal} {Nature}\ }%
  \textbf{\bibinfo {volume} {351}},\ \bibinfo {pages} {111} (\bibinfo {year}
  {1991})%
  \bibAnnoteFile{NoStop}{Scully1991}%
\bibitem{Herzog1995}%
  \BibitemOpen
  \bibfield{author}{%
  \bibinfo {author} {\bibfnamefont{T.~J.}\ \bibnamefont{Herzog}}, \bibinfo
  {author} {\bibfnamefont{P.~G.}\ \bibnamefont{Kwiat}}, \bibinfo {author}
  {\bibfnamefont{H.}~\bibnamefont{Weinfurter}},\ and\ \bibinfo {author}
  {\bibfnamefont{A.}~\bibnamefont{Zeilinger}},\ }%
  \bibfield{journal}{%
  \bibinfo {journal} {Phys. Rev. Lett.}\ }%
  \textbf{\bibinfo {volume} {75}},\ \bibinfo {pages} {3034} (\bibinfo {year}
  {1995})%
  \bibAnnoteFile{NoStop}{Herzog1995}%
\bibitem{Mir2007}%
  \BibitemOpen
  \bibfield{author}{%
  \bibinfo {author} {\bibfnamefont{R.}~\bibnamefont{Mir}}, \bibinfo {author}
  {\bibfnamefont{J.~S.}\ \bibnamefont{Lundeen}}, \bibinfo {author}
  {\bibfnamefont{M.~W.}\ \bibnamefont{Mitchell}}, \bibinfo {author}
  {\bibfnamefont{A.~M.}\ \bibnamefont{Steinberg}}, \bibinfo {author}
  {\bibfnamefont{J.~L.}\ \bibnamefont{Garretson}},\ and\ \bibinfo {author}
  {\bibfnamefont{H.~M.}\ \bibnamefont{Wiseman}},\ }%
  \bibfield{journal}{%
  \bibinfo {journal} {New J. Phys.}\ }%
  \textbf{\bibinfo {volume} {9}},\ \bibinfo {pages} {287} (\bibinfo {year}
  {2007})%
  \bibAnnoteFile{NoStop}{Mir2007}%
\bibitem{Hilmer2007}%
  \BibitemOpen
  \bibfield{author}{%
  \bibinfo {author} {\bibfnamefont{R.}~\bibnamefont{Hilmer}}\ and\ \bibinfo
  {author} {\bibfnamefont{P.}~\bibnamefont{Kwiat}},\ }%
  \bibfield{journal}{%
  \bibinfo {journal} {Sci. Am.}\ }%
  \textbf{\bibinfo {volume} {296}},\ \bibinfo {pages} {90} (\bibinfo {year}
  {2007})%
  \bibAnnoteFile{NoStop}{Hilmer2007}%
\bibitem{Wootters1998}%
  \BibitemOpen
  \bibfield{author}{%
  \bibinfo {author} {\bibfnamefont{W.~K.}\ \bibnamefont{Wootters}},\ }%
  \bibfield{journal}{%
  \bibinfo {journal} {Phys. Rev. Lett.}\ }%
  \textbf{\bibinfo {volume} {80}},\ \bibinfo {pages} {2245} (\bibinfo {year}
  {1998})%
  \bibAnnoteFile{NoStop}{Wootters1998}%
\bibitem{Samuelsson2005}%
  \BibitemOpen
  \bibfield{author}{%
  \bibinfo {author} {\bibfnamefont{P.}~\bibnamefont{Samuelsson}}, \bibinfo
  {author} {\bibfnamefont{E.~V.}\ \bibnamefont{Sukhorukov}},\ and\ \bibinfo
  {author} {\bibfnamefont{M.}~\bibnamefont{B\"uttiker}},\ }%
  \bibfield{journal}{%
  \bibinfo {journal} {New J. Phys.}\ }%
  \textbf{\bibinfo {volume} {7}},\ \bibinfo {pages} {176} (\bibinfo {year}
  {2005})%
  \bibAnnoteFile{NoStop}{Samuelsson2005}%
\bibitem{Frustaglia2009}%
  \BibitemOpen
  \bibfield{author}{%
  \bibinfo {author} {\bibfnamefont{D.}~\bibnamefont{Frustaglia}}\ and\ \bibinfo
  {author} {\bibfnamefont{A.}~\bibnamefont{Cabello}},\ }%
  \bibfield{journal}{%
  \bibinfo {journal} {Phys. Rev. B}\ }%
  \textbf{\bibinfo {volume} {80}},\ \bibinfo {pages} {201312} (\bibinfo {year}
  {2009})%
  \bibAnnoteFile{NoStop}{Frustaglia2009}%
\bibitem{Buks1998}%
  \BibitemOpen
  \bibfield{author}{%
  \bibinfo {author} {\bibfnamefont{E.}~\bibnamefont{Buks}}, \bibinfo {author}
  {\bibfnamefont{R.}~\bibnamefont{Schuster}}, \bibinfo {author}
  {\bibfnamefont{M.}~\bibnamefont{Heiblum}}, \bibinfo {author}
  {\bibfnamefont{D.}~\bibnamefont{Mahalu}},\ and\ \bibinfo {author}
  {\bibfnamefont{V.}~\bibnamefont{Umansky}},\ }%
  \bibfield{journal}{%
  \bibinfo {journal} {Nature}\ }%
  \textbf{\bibinfo {volume} {391}},\ \bibinfo {pages} {871} (\bibinfo {year}
  {1998})%
  \bibAnnoteFile{NoStop}{Buks1998}%
\bibitem{Sprinzak2000}%
  \BibitemOpen
  \bibfield{author}{%
  \bibinfo {author} {\bibfnamefont{D.}~\bibnamefont{Sprinzak}}, \bibinfo
  {author} {\bibfnamefont{E.}~\bibnamefont{Buks}}, \bibinfo {author}
  {\bibfnamefont{M.}~\bibnamefont{Heiblum}},\ and\ \bibinfo {author}
  {\bibfnamefont{H.}~\bibnamefont{Shtrikman}},\ }%
  \bibfield{journal}{%
  \bibinfo {journal} {Phys. Rev. Lett.}\ }%
  \textbf{\bibinfo {volume} {84}},\ \bibinfo {pages} {5820 } (\bibinfo {year}
  {2000})%
  \bibAnnoteFile{NoStop}{Sprinzak2000}%
\bibitem{Neder2007c}%
  \BibitemOpen
  \bibfield{author}{%
  \bibinfo {author} {\bibfnamefont{I.}~\bibnamefont{Neder}}, \bibinfo {author}
  {\bibfnamefont{F.}~\bibnamefont{Marquardt}}, \bibinfo {author}
  {\bibfnamefont{M.}~\bibnamefont{Heiblum}}, \bibinfo {author}
  {\bibfnamefont{D.}~\bibnamefont{Mahalu}},\ and\ \bibinfo {author}
  {\bibfnamefont{V.}~\bibnamefont{Umansky}},\ }%
  \bibfield{journal}{%
  \bibinfo {journal} {Nature Phys.}\ }%
  \textbf{\bibinfo {volume} {3}},\ \bibinfo {pages} {524} (\bibinfo {year}
  {2007})%
  \bibAnnoteFile{NoStop}{Neder2007c}%
\bibitem{Roulleau2009}%
  \BibitemOpen
  \bibfield{author}{%
  \bibinfo {author} {\bibfnamefont{P.}~\bibnamefont{Roulleau}}, \bibinfo
  {author} {\bibfnamefont{F.}~\bibnamefont{Portier}}, \bibinfo {author}
  {\bibfnamefont{P.}~\bibnamefont{Roche}}, \bibinfo {author}
  {\bibfnamefont{A.}~\bibnamefont{Cavanna}}, \bibinfo {author}
  {\bibfnamefont{G.}~\bibnamefont{Faini}}, \bibinfo {author}
  {\bibfnamefont{U.}~\bibnamefont{Gennser}},\ and\ \bibinfo {author}
  {\bibfnamefont{D.}~\bibnamefont{Mailly}},\ }%
  \bibfield{journal}{%
  \bibinfo {journal} {Phys. Rev. Lett.}\ }%
  \textbf{\bibinfo {volume} {102}},\ \bibinfo {pages} {236802} (\bibinfo {year}
  {2009})%
  \bibAnnoteFile{NoStop}{Roulleau2009}%
\bibitem{Dressel2010}%
  \BibitemOpen
  \bibfield{author}{%
  \bibinfo {author} {\bibfnamefont{J.}~\bibnamefont{Dressel}}, \bibinfo
  {author} {\bibfnamefont{S.}~\bibnamefont{Agarwal}},\ and\ \bibinfo {author}
  {\bibfnamefont{A.~N.}\ \bibnamefont{Jordan}},\ }%
  \bibfield{journal}{%
  \bibinfo {journal} {Phys. Rev. Lett.}\ }%
  \textbf{\bibinfo {volume} {104}},\ \bibinfo {pages} {240401} (\bibinfo {year}
  {2010})%
  \bibAnnoteFile{NoStop}{Dressel2010}%
\bibitem{Dressel2011}%
  \BibitemOpen
  \bibfield{author}{%
  \bibinfo {author} {\bibfnamefont{J.}~\bibnamefont{Dressel}}, \bibinfo
  {author} {\bibfnamefont{C.~J.}\ \bibnamefont{Broadbent}}, \bibinfo {author}
  {\bibfnamefont{J.~C.}\ \bibnamefont{Howell}},\ and\ \bibinfo {author}
  {\bibfnamefont{A.~N.}\ \bibnamefont{Jordan}},\ }%
  \bibfield{journal}{%
  \bibinfo {journal} {Phys. Rev. Lett.}\ }%
  \textbf{\bibinfo {volume} {106}},\ \bibinfo {pages} {040402} (\bibinfo {year}
  {2011})%
  \bibAnnoteFile{NoStop}{Dressel2011}%
\bibitem{Braginski1992}%
  \BibitemOpen
  \bibfield{author}{%
  \bibinfo {author} {\bibfnamefont{V.}~\bibnamefont{Braginski}}\ and\ \bibinfo
  {author} {\bibfnamefont{F.}~\bibnamefont{Khalili}},\ }%
  \emph{\bibinfo {title} {Quantum Measurement}}\ (\bibinfo {publisher}
  {Cambridge University Press},\ \bibinfo {year} {1992})%
  \bibAnnoteFile{NoStop}{Braginski1992}%
\bibitem{Nielsen2000}%
  \BibitemOpen
  \bibfield{author}{%
  \bibinfo {author} {\bibfnamefont{M.~A.}\ \bibnamefont{Nielsen}}\ and\
  \bibinfo {author} {\bibfnamefont{I.~L.}\ \bibnamefont{Chuang}},\ }%
  \emph{\bibinfo {title} {Quantum computation and quantum information}}\
  (\bibinfo {publisher} {Cambridge University Press},\ \bibinfo {year} {2000})%
  \bibAnnoteFile{NoStop}{Nielsen2000}%
\bibitem{Wiseman2009}%
  \BibitemOpen
  \bibfield{author}{%
  \bibinfo {author} {\bibfnamefont{H.~M.}\ \bibnamefont{Wiseman}}\ and\
  \bibinfo {author} {\bibfnamefont{G.}~\bibnamefont{Milburn}},\ }%
  \emph{\bibinfo {title} {Quantum Measurement and Control}}\ (\bibinfo
  {publisher} {Cambridge University Press},\ \bibinfo {year} {2009})%
  \bibAnnoteFile{NoStop}{Wiseman2009}%
\bibitem{Buttiker2002}%
  \BibitemOpen
  \bibfield{author}{%
  \bibinfo {author} {\bibfnamefont{S.}~\bibnamefont{Pilgram}}\ and\ \bibinfo
  {author} {\bibfnamefont{M.}~\bibnamefont{B\"uttiker}},\ }%
  \bibfield{journal}{%
  \bibinfo {journal} {Phys. Rev. Lett.}\ }%
  \textbf{\bibinfo {volume} {89}},\ \bibinfo {pages} {200401} (\bibinfo {year}
  {2002})%
  \bibAnnoteFile{NoStop}{Buttiker2002}%
\bibitem{Averin2005}%
  \BibitemOpen
  \bibfield{author}{%
  \bibinfo {author} {\bibfnamefont{D.~V.}\ \bibnamefont{Averin}}\ and\ \bibinfo
  {author} {\bibfnamefont{E.~V.}\ \bibnamefont{Sukhorukov}},\ }%
  \bibfield{journal}{%
  \bibinfo {journal} {Phys. Rev. Lett.}\ }%
  \textbf{\bibinfo {volume} {95}},\ \bibinfo {pages} {126803} (\bibinfo {year}
  {2005})%
  \bibAnnoteFile{NoStop}{Averin2005}%
\bibitem{Sukhorukov2007b}%
  \BibitemOpen
  \bibfield{author}{%
  \bibinfo {author} {\bibfnamefont{E.~V.}\ \bibnamefont{Sukhorukov}}, \bibinfo
  {author} {\bibfnamefont{A.~N.}\ \bibnamefont{Jordan}}, \bibinfo {author}
  {\bibfnamefont{S.}~\bibnamefont{Gustavsson}}, \bibinfo {author}
  {\bibfnamefont{R.}~\bibnamefont{Leturcq}}, \bibinfo {author}
  {\bibfnamefont{T.}~\bibnamefont{Ihn}},\ and\ \bibinfo {author}
  {\bibfnamefont{K.}~\bibnamefont{Ensslin}},\ }%
  \bibfield{journal}{%
  \bibinfo {journal} {Nat. Phys.}\ }%
  \textbf{\bibinfo {volume} {3}},\ \bibinfo {pages} {243} (\bibinfo {year}
  {2007})%
  \bibAnnoteFile{NoStop}{Sukhorukov2007b}%
\bibitem{Aharonov1988}%
  \BibitemOpen
  \bibfield{author}{%
  \bibinfo {author} {\bibfnamefont{Y.}~\bibnamefont{Aharonov}}, \bibinfo
  {author} {\bibfnamefont{D.~Z.}\ \bibnamefont{Albert}},\ and\ \bibinfo
  {author} {\bibfnamefont{L.}~\bibnamefont{Vaidman}},\ }%
  \bibfield{journal}{%
  \bibinfo {journal} {Phys. Rev. Lett.}\ }%
  \textbf{\bibinfo {volume} {60}},\ \bibinfo {pages} {1351 } (\bibinfo {year}
  {1988})%
  \bibAnnoteFile{NoStop}{Aharonov1988}%
\bibitem{Ritchie1991}%
  \BibitemOpen
  \bibfield{author}{%
  \bibinfo {author} {\bibfnamefont{N.~W.~M.}\ \bibnamefont{Ritchie}}, \bibinfo
  {author} {\bibfnamefont{J.~G.}\ \bibnamefont{Story}},\ and\ \bibinfo {author}
  {\bibfnamefont{R.~G.}\ \bibnamefont{Hulet}},\ }%
  \bibfield{journal}{%
  \bibinfo {journal} {Phys. Rev. Lett.}\ }%
  \textbf{\bibinfo {volume} {66}},\ \bibinfo {pages} {1107 } (\bibinfo {year}
  {1991})%
  \bibAnnoteFile{NoStop}{Ritchie1991}%
\bibitem{Pryde2005}%
  \BibitemOpen
  \bibfield{author}{%
  \bibinfo {author} {\bibfnamefont{G.~J.}\ \bibnamefont{Pryde}}, \bibinfo
  {author} {\bibfnamefont{J.~L.}\ \bibnamefont{O'Brien}}, \bibinfo {author}
  {\bibfnamefont{A.~G.}\ \bibnamefont{White}}, \bibinfo {author}
  {\bibfnamefont{T.~C.}\ \bibnamefont{Ralph}},\ and\ \bibinfo {author}
  {\bibfnamefont{H.~M.}\ \bibnamefont{Wiseman}},\ }%
  \bibfield{journal}{%
  \bibinfo {journal} {Phys. Rev. Lett.}\ }%
  \textbf{\bibinfo {volume} {94}},\ \bibinfo {pages} {220405} (\bibinfo {year}
  {2005})%
  \bibAnnoteFile{NoStop}{Pryde2005}%
\bibitem{Aharonov2008}%
  \BibitemOpen
  \bibfield{author}{%
  \bibinfo {author} {\bibfnamefont{Y.}~\bibnamefont{Aharonov}}\ and\ \bibinfo
  {author} {\bibfnamefont{L.}~\bibnamefont{Vaidman}},\ }%
  \bibfield{journal}{%
  \bibinfo {journal} {Lect. Notes Phys.}\ }%
  \textbf{\bibinfo {volume} {734}},\ \bibinfo {pages} {399 } (\bibinfo {year}
  {2008})%
  \bibAnnoteFile{NoStop}{Aharonov2008}%
\bibitem{Hosten2008}%
  \BibitemOpen
  \bibfield{author}{%
  \bibinfo {author} {\bibfnamefont{O.}~\bibnamefont{Hosten}}\ and\ \bibinfo
  {author} {\bibfnamefont{P.}~\bibnamefont{Kwiat}},\ }%
  \bibfield{journal}{%
  \bibinfo {journal} {Science}\ }%
  \textbf{\bibinfo {volume} {319}},\ \bibinfo {pages} {787} (\bibinfo {year}
  {2008})%
  \bibAnnoteFile{NoStop}{Hosten2008}%
\bibitem{Dixon2009}%
  \BibitemOpen
  \bibfield{author}{%
  \bibinfo {author} {\bibfnamefont{P.~B.}\ \bibnamefont{Dixon}}, \bibinfo
  {author} {\bibfnamefont{D.~J.}\ \bibnamefont{Starling}}, \bibinfo {author}
  {\bibfnamefont{A.~N.}\ \bibnamefont{Jordan}},\ and\ \bibinfo {author}
  {\bibfnamefont{J.~C.}\ \bibnamefont{Howell}},\ }%
  \bibfield{journal}{%
  \bibinfo {journal} {Phys. Rev. Lett.}\ }%
  \textbf{\bibinfo {volume} {102}},\ \bibinfo {pages} {173601} (\bibinfo {year}
  {2009})%
  \bibAnnoteFile{NoStop}{Dixon2009}%
\bibitem{Starling2009}%
  \BibitemOpen
  \bibfield{author}{%
  \bibinfo {author} {\bibfnamefont{D.~J.}\ \bibnamefont{Starling}}, \bibinfo
  {author} {\bibfnamefont{P.~B.}\ \bibnamefont{Dixon}}, \bibinfo {author}
  {\bibfnamefont{A.~N.}\ \bibnamefont{Jordan}},\ and\ \bibinfo {author}
  {\bibfnamefont{J.~C.}\ \bibnamefont{Howell}},\ }%
  \bibfield{journal}{%
  \bibinfo {journal} {Phys. Rev. A}\ }%
  \textbf{\bibinfo {volume} {80}},\ \bibinfo {pages} {041803} (\bibinfo {year}
  {2009})%
  \bibAnnoteFile{NoStop}{Starling2009}%
\bibitem{Howell2010}%
  \BibitemOpen
  \bibfield{author}{%
  \bibinfo {author} {\bibfnamefont{J.~C.}\ \bibnamefont{Howell}}, \bibinfo
  {author} {\bibfnamefont{D.~J.}\ \bibnamefont{Starling}}, \bibinfo {author}
  {\bibfnamefont{P.~B.}\ \bibnamefont{Dixon}}, \bibinfo {author}
  {\bibfnamefont{P.~K.}\ \bibnamefont{Vudyasetu}},\ and\ \bibinfo {author}
  {\bibfnamefont{A.~N.}\ \bibnamefont{Jordan}},\ }%
  \bibfield{journal}{%
  \bibinfo {journal} {Phys. Rev. A}\ }%
  \textbf{\bibinfo {volume} {81}},\ \bibinfo {pages} {033813} (\bibinfo {year}
  {2010})%
  \bibAnnoteFile{NoStop}{Howell2010}%
\bibitem{Zilberberg2011}%
  \BibitemOpen
  \bibfield{author}{%
  \bibinfo {author} {\bibfnamefont{O.}~\bibnamefont{Zilberberg}}, \bibinfo
  {author} {\bibfnamefont{A.}~\bibnamefont{Romito}},\ and\ \bibinfo {author}
  {\bibfnamefont{Y.}~\bibnamefont{Gefen}},\ }%
  \bibfield{journal}{%
  \bibinfo {journal} {Phys. Rev. Lett.}\ }%
  \textbf{\bibinfo {volume} {106}},\ \bibinfo {pages} {080405} (\bibinfo {year}
  {2011})%
  \bibAnnoteFile{NoStop}{Zilberberg2011}%
\bibitem{Leggett1985}%
  \BibitemOpen
  \bibfield{author}{%
  \bibinfo {author} {\bibfnamefont{A.~J.}\ \bibnamefont{Leggett}}\ and\
  \bibinfo {author} {\bibfnamefont{A.}~\bibnamefont{Garg}},\ }%
  \bibfield{journal}{%
  \bibinfo {journal} {Phys. Rev. Lett.}\ }%
  \textbf{\bibinfo {volume} {54}},\ \bibinfo {pages} {857 } (\bibinfo {year}
  {1985})%
  \bibAnnoteFile{NoStop}{Leggett1985}%
\bibitem{Williams2008}%
  \BibitemOpen
  \bibfield{author}{%
  \bibinfo {author} {\bibfnamefont{N.~S.}\ \bibnamefont{Williams}}\ and\
  \bibinfo {author} {\bibfnamefont{A.~N.}\ \bibnamefont{Jordan}},\ }%
  \bibfield{journal}{%
  \bibinfo {journal} {Phys. Rev. Lett.}\ }%
  \textbf{\bibinfo {volume} {100}},\ \bibinfo {pages} {26804} (\bibinfo {year}
  {2008})%
  \bibAnnoteFile{NoStop}{Williams2008}%
\bibitem{Tollaksen2007}%
  \BibitemOpen
  \bibfield{author}{%
  \bibinfo {author} {\bibfnamefont{J.}~\bibnamefont{Tollaksen}},\ }%
  \bibfield{journal}{%
  \bibinfo {journal} {J. Phys. A}\ }%
  \textbf{\bibinfo {volume} {40}},\ \bibinfo {pages} {9033} (\bibinfo {year}
  {2007})%
  \bibAnnoteFile{NoStop}{Tollaksen2007}%
\end{thebibliography}

%

\end{document}